\documentclass[12pt]{article}

\usepackage[left=2.8cm,right=2.8cm,top=2.8cm,bottom=2.8cm]{geometry}

\usepackage[utf8]{inputenc}

\usepackage{mathrsfs,mathtools,bbold}
\usepackage{amsmath,amssymb}
\usepackage{amsfonts}
\usepackage{verbatim}
\usepackage{graphicx}
\usepackage{cite}
\usepackage{mathrsfs,mathtools}
\usepackage{ccaption}
\usepackage[dvipsnames]{xcolor}
\definecolor{SchoolColor}{rgb}{0.6471, 0.1098, 0.1882} 
\usepackage{stmaryrd}
\usepackage{textcomp}
\usepackage[colorlinks=true,linkcolor=black,citecolor=teal,urlcolor=MidnightBlue,filecolor=black]{hyperref}
\usepackage{eucal} 

\newcommand{\unity}{1\hspace{-0.243em}\text{l}}

 \csname
@addtoreset\endcsname{equation}{section}

\DeclareMathOperator{\extdm}{d}
\newcommand{\extd}{\extdm \!}

\providecommand{\Lt}{{\tt L}}
\renewcommand{\Lt}{{\tt L}}

\providecommand{\Jt}{{\tt J}}
\renewcommand{\Jt}{{\tt J}}

\providecommand{\St}{{\tt S}}
\renewcommand{\St}{{\tt S}}

\providecommand{\Kt}{{\tt K}}
\renewcommand{\Kt}{{\tt K}}
\providecommand{\St}{{\tt S}}
\renewcommand{\St}{{\tt S}}

\begin{document}


\vspace{30pt}

\begin{center}


{\Large\sc Warped Black Holes in Lower-Spin Gravity}\\

---------------------------------------------------------------------------------------------------------


\vspace{25pt}
{\sc T.~Azeyanagi, S.~Detournay and M.~Riegler}
\vspace{10pt}

\sl\small
Physique Th{\'e}orique et Math{\'e}matique\\
Universit{\'e} libre de Bruxelles and International Solvay Institutes\\
Campus Plaine C.P. 231\\
B-1050 Bruxelles, Belgium
\vspace{10pt}

{\it tatsuo.azeyanagi@ulb.ac.be,\\ sdetourn@ulb.ac.be, \\ max.riegler@ulb.ac.be}

\vspace{50pt} {\sc\large Abstract} \end{center}

\noindent
We provide a simple holographic description for a warped conformal field theory (WCFT) at finite temperature. To this end we study the counterpart of warped anti-de Sitter black holes in three dimensions using a lower-spin $\mathfrak{sl}(2,\mathbb{R})\oplus\mathfrak{u}(1)$ Chern-Simons theory proposed by Hofman and Rollier. We determine the asymptotic symmetries, thermal entropy and holographic entanglement entropy and show that all these quantities are in perfect agreement with the expectations from the dual WCFT perspective. In addition we provide a metric interpretation of our results which naturally fits with our analysis in the Chern-Simons formulation.


\newpage


\tableofcontents
\hypersetup{linkcolor=SchoolColor}
\newpage

\section{Introduction}

For the largest part of their existences, anti-de Sitter (AdS) spaces and conformal field theories (CFTs) have followed fairly independent and lonesome paths. Anti-de Sitter spaces have been around since almost the advent of general relativity \cite{dS,dS2}. The origins of CFTs can be traced back to statistical mechanics, where they were identified as describing critical phenomena. Later, their importance was recognized in string theory, around at the same time the modern study of two-dimensional CFTs was initiated \cite{Belavin:1984vu}. Signs of interactions (apart from general isometry groups considerations) appeared in 1986 in the work of Brown and Henneaux \cite{Brown:1986nw} demonstrating that the asymptotic symmetries of pure AdS$_3$ gravity consisted in the two-dimensional conformal algebra. This important step establishing a link between AdS and CFT led to unexpected breakthroughs in gravitational physics, such as, if one had to name only one, the beautiful interplay between two-dimensional CFTs and black hole entropy, which crystallizes in the famous Cardy formula\cite{Cardy:1986ie,Bloete:1986qm,Strominger:1997eq, Carlip:2000nv} and eventually culminates in the AdS/CFT correspondence \cite{Maldacena:1997re,Gubser:1998bc,Witten:1998qj}.

Over the last years, an important collective effort has been devoted in extending the principles of AdS/CFT to more general setups -- most notably to non-AdS backgrounds on the gravity side. This is because the holographic nature of gravity in general, and the Bekenstein-Hawking area law in particular are not supposed to be contingent to AdS spaces (see e.g. \cite{Strominger:1996sh}). A notorious proposal in the context is the Kerr/CFT correspondence \cite{Guica:2008mu}, suggesting a holographic duality involving (the near-horizon region of) extremal four-dimensional Kerr black holes (see also \cite{Castro:2009jf} for near-extremal BHs). The relevant geometry is the so-called Bardeen-Horowitz metric (or NHEK geometry) with SL$(2,\mathbb{R})\times$U$(1)$ isometry \cite{Bardeen:1999px}, which turns out to be universal \cite{Kunduri:2013ana}, but importantly does not include the generic AdS$_3$ factor familiar from the near-horizon geometry of supersymmetric black holes and allowing us to apply AdS/CFT techniques. At fixed polar angle, the NHEK geometry reduces to a three-dimensional metric called (self-dual spacelike) warped AdS$_3$ (WAdS$_3$). Geometrically, they can be understood as a Kerr-Schild-like deformation of AdS$_3$ using a chiral Killing vector of its SO(2,2) isometry group (see e.g. \cite{Gurses:1994bjn, Nutku:1993eb,Israel:2004vv}. Depending on the type of the latter, one obtains timelike, spacelike and null warped AdS$_3$ spaces. The former can be identified with G{\"o}del space \cite{Rooman:1998xf}, while performing identifications in the latter two yields a variety of black hole solutions \cite{Bouchareb:2007yx, Anninos:2008fx, Anninos:2008qb} sharing similarities with the Ba{\~n}ados-Teitelboim-Zanelli (BTZ) black holes \cite{Banados:1992wn, Banados:1992gq} of AdS$_3$ gravity, and reducing to them when the deformation goes to zero. The asymptotic behavior of WAdS$_3$ spaces differs drastically from that of AdS$_3$ (in particular, they do not satisfy Brown-Henneaux boundary conditions), so WAdS$_3$ black holes are often viewed as a prototype for non-AdS black hole holography, possibly allowing us to get insights into (the near-horizon geometry of) their higher-dimensional cousin.

The departure from a usual AdS/CFT scenario for WAdS$_3$ spaces is crucially reflected in their asymptotic symmetries, consisting in the semidirect product of a Virasoro algebra and an affine $\hat{\mathfrak{u}}(1)$ algebra instead of the full two-dimensional conformal algebra \cite{Compere:2007in}.\footnote{Note that the full conformal algebra can in some embeddings be recovered when extra matter fields are present \cite{Guica:2013jza}.} This observation has been taken as a starting point for the study of holographic properties of WAdS$_3$, in particular through that of two-dimensional field theories with the corresponding symmetries. This led to the definition of a new type of field theories: warped conformal field theories (WCFTs).  These are two-dimensional field theories breaking Lorentz symmetry and possessing an infinite number of conserved charges satisfying a Virasoro-Kac-Moody $\hat{\mathfrak{u}}(1)$ algebra. Hofman and Strominger \cite{Hofman:2011zj} showed that under certain generic assumptions, a two-dimensional field theory with translation and chiral scale invariance is either a CFT or a WCFT (this is a warped version of the result of Polchinski that scale invariance implies conformal invariance \cite{Polchinski:1987dy}). It is interesting to observe that, contrary to what happened in the unfolding of AdS/CFT (where both sides had been independently known but did not talk to each other), here gravity suggested the existence of a new class of integrable field theories that had not been encountered so far, of which the mere existence was not even guaranteed a priori. Since then however, many steps have been taken towards defining and analyzing the properties of WCFTs: derive a Cardy-type formula and matching with black hole entropy \cite{Detournay:2012pc}, finding explicit examples and calculating partition functions \cite{Castro:2015uaa}, studying phase transitions \cite{Detournay:2015ysa} as well as correlation functions \cite{Song:2017czq}, matching of one-loop determinants \cite{Castro:2017mfj}, calculating entanglement entropy \cite{Castro:2015csg, Song:2016gtd} and the study of anomalies \cite{Jensen:2017tnb}.

Facing this new challenge, a natural approach would be to set up the simplest holographic toy-model model capturing the properties of a WCFT and see how far one can get. What we are looking for is the warped counterpart of pure Einstein-Hilbert gravity with a negative cosmological constant for three-dimensional gravity and AdS/CFT. Thanks to the absence of degrees of freedom and the presence of BTZ black holes in its spectrum, pure three-dimensional gravity might be the best candidate for a solvable model with quantum black holes. The latter theory has been shown to exhibit an extremely rich structure \cite{Witten:2007kt, Maloney:2007ud}. WAdS$_3$ spaces, on the other hand, are not Einstein spaces. For that reason, the holographic models considered so far for WCFTs either consist of higher-curvature gravity theories (see e.g.\cite{Compere:2007in,Compere:2008cv,Blagojevic:2008bn,Compere:2009zj,Blagojevic:2009ek,Henneaux:2011hv,Donnay:2015iia,Ghodsi:2010gk,Ghodsi:2010ev,Ghodsi:2011ua,Detournay:2016gao}) or require couplings to matter(see e.g. \cite{Israel:2004vv, Compere:2008cw, Levi:2009az,  Detournay:2010rh, ElShowk:2011cm, Song:2011sr, Guica:2011ia, Detournay:2012dz,Azeyanagi:2012zd, Guica:2013jza, Jeong:2014iva, Colgain:2015ela}). In any of the cases, the models possess local degrees of freedom. A noticeable exception is the model, dubbed lower-spin gravity proposed in \cite{Hofman:2014loa}, that can be described using a SL$(2,\mathbb{R})\times$U$(1)$ Chern-Simons theory. It was argued that this model is the minimal setup for the holographic description of WCFTs, much like the SL$(2,\mathbb{R})\times$SL$(2,\mathbb{R})$ Chern-Simons theory, which is classically equivalent to the Einstein-Hilbert action \cite{Witten:1988hc}, is the minimal setup for two-dimensional CFTs. In \cite{Hofman:2014loa}, a dictionary was provided for translating Chern-Simons gauge fields into a metriclike object. By using this, the authors demonstrated how vacuum WAdS$_3$ spacetimes can be encoded in the gauge fields. However, black hole solutions on WAdS$_3$ have not been discussed at all. 
In this paper, we will be concerned with the study of how  
spacelike WAdS$_3$ black holes are encoded in the SL$(2,\mathbb{R})\times$U$(1)$ Chern-Simons theory. 
We will explicitly construct the configurations of the 
Chern-Simons gauge fields corresponding to these black holes
and then explain how these configurations can be identified 
as black holes carrying nonzero entropy. 
This is the first step towards understanding of
thermodynamic properties of this theory through a holographic duality.

In AdS$_3$/CFT$_2$, one can essentially indistinctly work either in the metric or in the Chern-Simons formulation, as long as one is concerned with semiclassical considerations \cite{Witten:2007kt}. Most of the quantities on one side can almost unambiguously be defined on the other side. For instance, determining the Hawking temperature can be done be requiring the absence of a conical singularity in the Euclidean metric close to the horizon, which in the Chern-Simons formulation amounts to requiring that the holonomy of the connection along a certain cycle is trivial. We will identify the gauge connection counterparts of WAdS$_3$ black holes, study their thermodynamics and compare to the predictions from WCFT.

Another motivation to study warped black hole solutions in lower-spin gravity is related to its similarity to higher-spin gravity theories in AdS$_3$ that can also be described in terms of a Chern-Simons connection with a specific gauge algebra (see e.g. \cite{Campoleoni:2010zq,Campoleoni:2011hg,Campoleoni:2012hp,Henneaux:2010xg}). For the higher-spin theories in AdS$_3$ this algebra is $\mathfrak{sl}(N,\mathbb{R})\oplus\mathfrak{sl}(N,\mathbb{R})$ \cite{Campoleoni:2010zq}. Since things such as an event horizon are not gauge invariant objects any more, as soon as higher-spin symmetries are present, one needs to find other ways to define thermodynamically sensible black holes with higher-spin charges. In the Chern-Simons formulation this is usually done by requiring that the holonomies of the gauge connection satisfy certain requirements (see e.g. \cite{Gutperle:2011kf,Ammon:2011nk}). These requirements are basically that the holonomies of the higher-spin connections have the same eigenvalues as the corresponding connection describing the BTZ black hole in AdS$_3$. Thus if one is interested in possible higher-spin extensions of warped black holes in a Chern-Simons formulation, one first needs to understand how to describe an ordinary warped black hole in this setup. Providing basic understanding of warped AdS$_3$ black holes in terms of a lower-spin Chern-Simons theory is another motivation for this work.

In order to describe the thermodynamics of spacelike warped AdS black holes in the Chern-Simons formulation, we will follow the strategy that is quite similar to the one used for describing black holes in higher-spin theories in AdS$_3$ (which itself is inspired by the analysis of the thermodynamics of the BTZ black hole in AdS$_3$ in Chern-Simons formulation \cite{Banados:1992wn,Banados:1992gq})\footnote{We provide a brief review of certain aspects of BTZ thermodynamics in the Chern-Simons formulation that are relevant for our work in Appendix~\ref{sec:AppendixA}.}:
    \begin{itemize}
        \item Mass and angular momentum are the canonical boundary charges that are associated to translations in a timelike and angular direction, respectively. Thus, we impose that the following relations between the Chern-Simons gauge parameters,  $\varepsilon$ for $\mathfrak{sl}(2,\mathbb{R})$ and  $\bar{\varepsilon}$ for $\mathfrak{u}(1)$, and associated Killing vectors $\xi^\mu$ holds,
        \begin{equation}
            \varepsilon=\xi^\mu\mathcal{A}_\mu,\qquad\bar{\varepsilon}=\xi^\mu\mathcal{C}_\mu,
        \end{equation}
        where $\mathcal{A}$ and $\mathcal{C}$, respectively are the $\mathfrak{sl}(2,\mathbb{R})$ and $\mathfrak{u}(1)$ Chern-Simons gauge fields.
        \item The inverse temperature $\beta$ and angular velocity $\Omega$ as functions of mass and angular momentum are determined by requiring that the radially independent parts of the Chern-Simons connections $\mathcal{A}$ and $\mathcal{C}$, denoted by $a_\varphi$, $a_t$, $c_\varphi$ and $c_t$, satisfy the following conditions:\footnote{See \eqref{eq:BTZHolonomiesLorentzian} for the conditions that have to be satisfied in the BTZ case and that inspired us to impose the conditions \eqref{eq:IntroHolonomyConditions}.}
        \begin{subequations}\label{eq:IntroHolonomyConditions}
        \begin{align}
		\textrm{Eigen}\left[h\right] & = \textrm{Eigen}\left[2\pi \Lt_0\right],\\
		\textrm{Eigen}\left[\bar{h}\right] & =\textrm{Eigen}\left[2\pi \gamma \St\right],
        \end{align}		
	\end{subequations}
	where $\textrm{Eigen}[...]$ denotes the eigenvalues of ... 
    \begin{subequations}
    \begin{align}
        h & =\frac{\beta}{2\pi}\left(\int\extd\varphi\,a_t + \Omega \int\extd\varphi\,a_\varphi\right),\\\
        \bar{h} & =\frac{\beta}{2\pi}\left(\int\extd\varphi\,c_t + \Omega \int\extd\varphi\,c_\varphi\right),
    \end{align}
    \end{subequations}
    $\gamma$ is an undetermined parameter (to be determined later) and $\Lt_0$ and $\St$, respectively are generators of $\mathfrak{sl}(2,\mathbb{R})$ and $\mathfrak{u}(1)$. Thus, in order to have sensible thermodynamics, we are requiring that the eigenvalues of $h$ coincide with the eigenvalues of $2\pi \Lt_0$ and similarly the eigenvalues of $\bar{h}$ coincide with the eigenvalues of $2\pi\gamma\St$. 
        \item We require that the vacuum solution of the spacelike warped AdS$_3$ black hole in this Chern-Simons formulation is defined by having a ``warped-trivial'' holonomy around the $\varphi$-cycle:
    \begin{equation}\label{eq:IntroCSVacuum}
        e^{\oint a_\varphi}=-\unity,\quad e^{\oint c_\varphi}=e^{2\pi i \gamma}.
    \end{equation}
    \end{itemize}
We will look in detail these points in the main body of the paper.

This paper is organized as follows. In Sec.~\ref{sec:CSWAdS3BH} we define our setup, calculate the asymptotic symmetry algebra and the thermal entropy in the Chern-Simons formulation. In Sec.~\ref{Sec:VacuumCSWCFT} we show how to define the vacuum state of our configuration in the Chern-Simons formulation and give supporting arguments for the requirement \eqref{eq:IntroHolonomyConditions} using WCFT arguments as well as \eqref{eq:IntroCSVacuum}. Section~\ref{Sec:HEEWilson} will be concerned with computing holographic entanglement entropy using Wilson lines. This computation also provides an independent check of the thermal entropy and thus, in turn, also the validity of the conditions \eqref{eq:IntroHolonomyConditions}. Section~\ref{Sec:Metric} will be focused on a metric interpretation of the Chern-Simons results.  This metric interpretation provides another explanation for the validity of all the requirements and, in addition, allows us to fix the previously undetermined parameter $\gamma$ in terms of geometric variables. Finally, the conclusion and outlook of this work are summarized in
Sec. \ref{conclusion_outlook}.
For comparison,
in Appendix \ref{sec:AppendixA}, we summarized how the usual BTZ black holes are described
in the metric formulation as well as in the 
Chern-Simons formulation with $\mathfrak{sl}(2, \mathbb{R})\oplus\mathfrak{sl}(2, \mathbb{R})$ gauge symmetry. 

\section{Spacelike WAdS$_3$ Black Holes in a Chern-Simons Formulation}\label{sec:CSWAdS3BH}

In order to describe spacelike warped AdS$_3$ black holes, we use a $\mathfrak{sl}(2,\mathbb{R})\oplus\mathfrak{u}(1)$ Chern-Simons formulation, very similar to the one presented in \cite{Hofman:2014loa}. In accordance with \cite{Hofman:2014loa}, we also call it lower-spin gravity. The action is given as follows: 
	\begin{equation}\label{eq:CSAction}
		I_{\textrm{CS}} = \frac{k}{4\pi} \int_{\mathcal{M}} \langle\mathcal{A} \wedge \extd \mathcal{A} +\frac23 \mathcal{A} \wedge \mathcal{A} \wedge \mathcal{A}\rangle+\frac{\kappa}{8\pi}\int_{\mathcal{M}}\left\langle \mathcal{C}\wedge\extd \mathcal{C}\right\rangle\,,
	\end{equation}
where $\langle\ldots\rangle$ is an appropriate invariant bilinear form, $k$ is the Chern-Simons coupling, $\kappa$ the $\mathfrak{u}(1)$ coupling and $\mathcal{M}$ a 2+1-dimensional manifold. The gauge field $\mathcal{A}$ takes values in $\mathfrak{sl}(2,\mathbb{R})$ and the second gauge field $\mathcal{C}$ in $\mathfrak{u}(1)$. 
We take the topology of the manifold $\mathcal{M}$ to be 
a cylinder with coordinates $0\leq \rho<\infty$, $-\infty<t<\infty$ and ${\varphi\in[0,2\pi]}$. Here $t$ is a temporal coordinate while
$\rho$ and $\varphi$ are spatial. Choosing the basis\footnote{A matrix representation of the $\mathfrak{sl}(2,\mathbb{R})$ part is given by
    	\begin{equation}
		\Lt_0 =\left(
			\begin{array}{ccc}
				\frac{1}{2}&0\\
				0&-\frac{1}{2}
			\end{array}\right)\,,\;
		\Lt_1 =\left(
			\begin{array}{ccc}
				0&0\\
				1&0
			\end{array}\right)\,,\;
		\Lt_{-1} =\left(
			\begin{array}{ccc}
				0&-1\\
				0&0
			\end{array}\right).\nonumber
	\end{equation}
} of $\mathfrak{sl}(2,\mathbb{R})$ ($\Lt_n$) and $\mathfrak{u}(1)$ ($\St$) generators as
\begin{equation}\label{eq:AdSxR:CommRel}
    [\Lt_n,\,\Lt_m]={}(n-m)\,\Lt_{n+m},\qquad[\Lt_n,\,\St]={}0,\qquad [\St,\,\St]={}0,
\end{equation}
the invariant bilinear form in \eqref{eq:CSAction} is given by
	\begin{equation}\label{eq:ToySLInvBilForm}
		\langle \Lt_n\Lt_m\rangle  =\left(
			\begin{array}{c|ccc}
				  &\Lt_1&\Lt_0&\Lt_{-1}\\
				\hline
				\Lt_1&0&0&-1\\
				\Lt_0&0&\frac{1}{2}&0\\
				\Lt_{-1}&-1&0&0
			\end{array}\right)\equiv\eta_{nm},\quad
		\langle \St\St\rangle  = 1.	
	\end{equation}
Some comments are in order regarding the way we 
present the gauge algebra here, especially the $\mathfrak{u}(1)$-part.	Introducing explicitly the generator $\St$ and the corresponding invariant bilinear form \eqref{eq:ToySLInvBilForm} might seem unnecessary at first sight. However, we stress that one of our motivations for this work is to provide a reference example of spacelike warped AdS$_3$ black holes in the Chern-Simons formulation that can be possibly extended to include higher-spin excitations.
The above choice is useful for this purpose. In the well-known AdS$_3$ higher-spin case, the embedding of the pure gravity sector into the higher-spin sector determines what kind of higher-spin fields are present in the resulting higher-spin theory (see e.g. \cite{Campoleoni:2011hg}). In a similar spirit we suggest that the field content of possible higher-spin extensions of spacelike warped AdS$_3$ black holes is determined by how the basic symmetries \eqref{eq:AdSxR:CommRel} are embedded into the higher-spin symmetries, $\mathfrak{sl}(2,\mathbb{R})\oplus\mathfrak{u}(1)\hookrightarrow\mathfrak{sl}(N,\mathbb{R})$. 

\subsection{Boundary conditions and asymptotic symmetries}\label{sec:BCsASA}

After the introduction of the basic setup for describing spacelike warped AdS$_3$ black holes in this work, the next step is to write down boundary conditions that include such black hole solutions. Since it has already been shown in \cite{Hofman:2014loa} that a model like \eqref{eq:CSAction} can describe spacelike, timelike and null warped AdS$_3$, we will take the boundary conditions presented in \cite{Hofman:2014loa} as an inspiration to write down suitable boundary conditions for a spacelike warped AdS$_3$ black hole\footnote{Later on we will argue that depending on the choice of chemical potentials the boundary conditions of \cite{Hofman:2014loa} even include spacelike warped AdS$_3$ black holes.}. 

First we will use some of the gauge freedom to fix the radial dependence of the gauge fields $\mathcal{A}$ and $\mathcal{C}$ as
	\begin{subequations}
	\begin{align}
		\mathcal{A}(\rho,t,\varphi) & =b^{-1}(\rho)\left[a(t,\varphi)+\extd\,\right]b(\rho),\\
		\mathcal{C}(\rho,t,\varphi)&=c(t,\varphi),
	\end{align}	
	\end{subequations}
with
	\begin{subequations}
	\begin{align}
		a(t,\varphi)&=a_\varphi(t,\varphi)\extd\varphi+a_t(t,\varphi)\extd t,\\
		c(t,\varphi)&=c_\varphi(t,\varphi)\extd\varphi+c_t(t,\varphi)\extd t.
	\end{align}	
	\end{subequations}
From the Chern-Simons perspective, the exact form of the group elements $b$ does not have any relevance for computing asymptotic symmetries as well as thermal properties of the physical system described by the Chern-Simons theory. The exact form of this group element, however, is important for the geometrical interpretation of the boundary conditions presented in this subsection. Since we will also present a metric interpretation in Sec.~\ref{Sec:Metric}, we fix the group element to be
	\begin{equation}
		b(\rho)=e^{\rho\Lt_0}.
	\end{equation}
There are in fact two reasons to choose this specific expression. The first reason is that this is a very common choice\footnote{There are also other possible choices such as in e.g. \cite{Grumiller:2016pqb}.} for AdS$_3$ gravity in the Chern-Simons formulation since the resulting metric takes the form of a Fefferman-Graham expansion. The second one is that this kind of gauge has also been used in \cite{Hofman:2014loa} and, thus, is helpful when making contact with the results in this work.

Using this gauge, we propose the following boundary conditions:
	\begin{subequations}\label{eq:WAdSBCs}
	\begin{align}
		a_\varphi&=\Lt_1-\mathfrak{L}\Lt_{-1},&
		a_t&=\mu \Lt_1+\omega_1 \Lt_0+\omega_2\Lt_{-1},\\
		c_\varphi&=\frac{4\pi}{\kappa}\mathcal{K}\St,& c_t&=\left(\nu+\frac{4\pi}{\kappa}\mathcal{K}\mu\right)\St,
	\end{align}
	\end{subequations}
Here $\mathfrak{L}:=\frac{2\pi}{k}\left(\mathcal{L}-\frac{2\pi}{\kappa}\mathcal{K}^2\right)$ and the functions $\mathcal{L}$, $\mathcal{K}$, $\mu$ and $\nu$ are in principle arbitrary functions of $t$ and $\varphi$. With a bit of hindsight, we interpret the functions $\mathcal{L}$ and $\mathcal{K}$ as functions characterizing the physical state and the functions $\mu$ and $\nu$ as chemical potentials.\footnote{The reason why the functions $\mathcal{L}$ and $\mathcal{K}$ characterize the physical state is that they appear in the canonical boundary charges. In addition, the equations of motion only fix the time evolution of the functions $\mathcal{L}$ and $\mathcal{K}$ but not of $\mu$ and $\nu$ thus specifying $\mathcal{L}$ and $\mathcal{K}$ as the dynamical variables.} This means in particular that we assume those chemical potentials to be fixed, i.e. $\delta\mu=\delta\nu=0$. The functions $\omega_{a}$ are fixed by the equations of motion, i.e. $\extd\mathcal{A}+[\mathcal{A},\mathcal{A}]=0$ and $\extd \mathcal{C}=0$.
For arbitrary but fixed chemical potentials, the equations of motion determine the time evolution of the state-dependent functions as well as $\omega_a$ as
    \begin{subequations}\label{eq:CSEOM}
    \begin{align}
        \partial_t\mathcal{L} & = \mu\mathcal{L}'+2\mathcal{L}\mu'-\frac{k}{4\pi}\mu'''+\mathcal{K}\nu',&
        \partial_t\mathcal{K} & = \mu\mathcal{K}'+\mathcal{K}\mu'+\frac{\kappa}{4\pi}\nu',\\
        \omega_1 & = -\mu',& \omega_2&=-\mathfrak{L}\mu+\frac{\mu''}{2},
    \end{align}
    \end{subequations}
where prime denotes a derivative with respect to $\varphi$. 

The next step is to find the gauge transformations that preserve the boundary conditions \eqref{eq:WAdSBCs}.
They are given by
    \begin{equation}
        \delta_\varepsilon\mathcal{A}_\mu=\partial_\mu\varepsilon+[\mathcal{A}_\mu,\varepsilon],\qquad\delta_{\bar{\varepsilon}}\mathcal{C}_\mu=\partial_\mu\bar{\varepsilon},
    \end{equation}
where    
	\begin{equation}\label{eq:GaugeParameter}
		\varepsilon(t,\varphi)=b^{-1}\left[\sum\limits_{a=-1}^1\epsilon^a(t,\varphi)\Lt_a\right]b,\qquad
		\bar{\varepsilon}(t,\varphi) = \epsilon^S(t,\varphi)\St,
	\end{equation}
with
	\begin{equation}
		\epsilon^1=\epsilon,\qquad \epsilon^0=-\epsilon',\qquad
		\epsilon^{-1}=-\mathfrak{L}\epsilon+\frac{\epsilon''}{2},\qquad
		\epsilon^S=\sigma+\frac{4\pi}{\kappa}\mathcal{K}\epsilon.	
	\end{equation}
These gauge transformations lead to the following infinitesimal transformation behavior of the functions $\mathcal{L}$ and $\mathcal{K}$:
	\begin{subequations}\label{eq:WAdSGaugeTrafos}
	\begin{align}
		\delta\mathcal{L}&=\epsilon\mathcal{L}'+2\mathcal{L}\epsilon'+\mathcal{K}\sigma'-\frac{k}{4\pi}\epsilon''',\\
		\delta\mathcal{K}&=\epsilon\mathcal{K}'+\mathcal{K}\epsilon'+\frac{\kappa}{4\pi}\sigma'.
	\end{align}
	\end{subequations}
In addition, the gauge parameters have to satisfy
    \begin{equation}
        \partial_t\epsilon=\mu\epsilon',\quad \partial_t\sigma=-\frac{4\pi}{\kappa}\mu\left(\epsilon\mathcal{K}\right)'-\epsilon\nu'.
    \end{equation}
Accordingly the variation of the canonical boundary charge is given by\footnote{For more details see e.g. \cite{Henneaux:1992,Blagojevic:2002aa}.}
	\begin{align}
		\delta Q[\varepsilon]+\delta Q[\bar{\varepsilon}] & =\frac{k}{2\pi}\int\extd\varphi\left\langle\varepsilon\,\delta\mathcal{A}_\varphi\right\rangle+\frac{\kappa}{4\pi}\int\extd\varphi\left\langle\bar{\varepsilon}\,\delta\mathcal{C}_\varphi\right\rangle\nonumber\\
		& =\int\extd\varphi\left(\delta\mathcal{L}\epsilon+\delta\mathcal{K}\sigma\right),
	\end{align}
which can be directly integrated to obtain the canonical boundary charge
	\begin{equation}\label{eq:WAdSCharge}
		Q=\int\extd\varphi\left(\mathcal{L}\epsilon+\mathcal{K}\sigma\right).
	\end{equation}
Using this canonical boundary charge as well as \eqref{eq:WAdSGaugeTrafos}, one can readily determine the following Dirac bracket algebra:
	\begin{subequations}
	\begin{align}
		\{\mathcal{L}(\varphi),\mathcal{L}(\bar{\varphi})\}&=2\mathcal{L}\delta'-\delta\mathcal{L}'-\frac{k}{4\pi}\delta''',\\
		\{\mathcal{L}(\varphi),\mathcal{K}(\bar{\varphi})\}&=\mathcal{K}\delta'-\delta\mathcal{K}',\\
		\{\mathcal{K}(\varphi),\mathcal{K}(\bar{\varphi})\}&=\frac{\kappa}{4\pi}\delta',
	\end{align}
	\end{subequations}
where all functions appearing on the r.h.s are functions of $\bar{\varphi}$ and prime denotes differentiation with respect to the corresponding argument. We have also defined ${\delta\equiv\delta(\varphi-\bar{\varphi})}$ and $\delta'\equiv\partial_\varphi\delta(\varphi-\bar{\varphi})$.

One can also expand the functions $\mathcal{L}$, $\mathcal{K}$ and delta functions in terms of Fourier modes as
	\begin{equation}\label{eq:FourierModes}
		\mathcal{L} =\frac{1}{2\pi}\sum\limits_{n\in\mathbb{Z}}\Lt_ne^{-in\varphi},\qquad
		\mathcal{K}=\frac{1}{2\pi}\sum\limits_{n\in\mathbb{Z}}\Kt_ne^{-in\varphi},\qquad
		\delta =\frac{1}{2\pi}\sum\limits_{n\in\mathbb{Z}}e^{-in(\varphi-\bar{\varphi})},
	\end{equation}
and then replace the Dirac brackets with commutators using $i\{\cdot,\cdot\}\rightarrow[\cdot,\cdot]$. In the end, we obtain the following commutation relations:
	\begin{subequations}\label{eq:WAdSASA}
	\begin{align}
		[\Lt_n,\,\Lt_m]&=(n-m)\,\Lt_{n+m}+\frac{c}{12}n(n^2-1)\delta_{n+m,0}\,,\\
		[\Lt_n,\,\Kt_m]&=-m\,\Kt_{n+m}\,,\\
		[\Kt_n,\,\Kt_m]&=\frac{\kappa}{2}\, n\,\delta_{n+m,0}\,,
	\end{align}
	\end{subequations}
with $c=6k$.
This algebra is a semidirect sum of a Virasoro algebra and an affine $\hat{\mathfrak{u}}(1)$ current algebra, matching with the basic symmetry for WCFTs. This also coincides with the asymptotic symmetry algebra found in \cite{Hofman:2014loa}. This is not a surprise, since the $\varphi$-part of the connections $\mathcal{A}$ and $\mathcal{C}$ coincide with that of the corresponding gauge fields in \cite{Hofman:2014loa}.

\subsection{Variational principle and holographic ward identities}\label{sec:VariationalPrinciple}

An important consistency check of the boundary conditions \eqref{eq:WAdSBCs} is to see whether or not they lead to a well defined variational principle.\footnote{We would like to thank an anonymous referee for suggesting to perform the additional checks found in this subsection.} Varying the action \eqref{eq:CSAction} one obtains on-shell
	\begin{equation}\label{eq:CSActionVar}
		\delta I_{\textrm{CS}} = \frac{k}{4\pi} \int_{\partial\mathcal{M}} \langle\delta\mathcal{A} \wedge \mathcal{A}\rangle+\frac{\kappa}{8\pi}\int_{\partial\mathcal{M}}\left\langle \delta\mathcal{C}\wedge\mathcal{C}\right\rangle\,.
	\end{equation}
It is straightforward to check that this term does not vanish for the boundary conditions \eqref{eq:WAdSBCs} with fixed chemical potentials and that one has to add an additional boundary term $I_\textrm{B}$ to the action \eqref{eq:CSAction}. This does not come as a surprise since the necessity of such a boundary term has already been discussed for a subset of our boundary conditions -- albeit using a slightly different notation -- in \cite{Hofman:2014loa}. In terms of the components of the gauge fields $a$ and $c$ that are independent of the radial coordinate $\rho$, the necessary boundary term is given by
    \begin{equation}\label{eq:BoundaryTerm}
        I_\textrm{B}=\frac{k}{4\pi}\int_{\partial\mathcal{M}}\langle (a_\varphi-2\Lt_1-\frac{\kappa}{2k}\langle c_\varphi^2\rangle\Lt_{-1})a_t\rangle-\frac{\kappa}{8\pi}\int_{\partial\mathcal{M}}\langle c_tc_\varphi\rangle.
    \end{equation}
It should be noted that the form of this boundary term closely resembles the form of the boundary terms encountered in Chern-Simons models of higher-spin AdS$_3$ theories, see e.g. \cite{Campoleoni:2010zq,deBoer:2013gz,deBoer:2014fra}. The variation of the total Chern-Simons action $I_{\textrm{tot}}=I_{\textrm{CS}}+I_{\textrm{B}}$ then vanishes for fixed chemical potentials, as expected.

In holographic setups where the gravity side can be described in terms of a Chern-Simons theory there is an intimate relation between the Ward identities of the dual quantum field theory and the flatness conditions of the Chern-Simons gauge field (see e.g. \cite{Banados:2004nr,Li:2015osa}). This relation is crucial for setting up the holographic dictionary between the functions $\mathcal{L}$ and $\mathcal{K}$ and the corresponding expectation values (EVs) in the dual WCFT. In \cite{Hofman:2014loa} it has been shown that this relation also extends to WCFTs that are described via lower-spin gravity theories i.e. the flatness conditions of the Chern-Simons gauge fields exactly reproduce the WCFT Ward identities. If in addition to that the variation of the total Chern-Simons action $I_{\textrm{CS}}$ takes the following schematic form on-shell
    \begin{equation}\label{eq:SchematicOnShellVariation}
        \delta I_\textrm{tot}\sim\int_{\partial\mathcal{M}}( \textrm{EVs})\delta(\textrm{sources}),
    \end{equation}
one obtains a functional that automatically solves the WCFT Ward identities and one can uniquely identify which functions in the Chern-Simons connection \eqref{eq:WAdSBCs} correspond to the expectation values of the dual WCFT currents and which functions correspond to the corresponding sources \cite{Banados:2004nr,deBoer:2014fra}. Thus, in order to set up this dictionary in the case at hand we will proceed with first deriving the WCFT Ward identities in the presence of sources\footnote{As far as we are aware of these Ward identities were not computed anywhere else before in the literature.} and then determine the exact form of \eqref{eq:SchematicOnShellVariation} for the boundary conditions \eqref{eq:WAdSBCs}.

Let us assume a two-dimensional WCFT with coordinates $x$ and $y$ such that by introducing $\varphi=x-y$ and $t=x+y$ the basic symmetries of the WCFT are \cite{Detournay:2012pc}
    \begin{equation}
        \varphi\rightarrow f(\varphi),\quad t\rightarrow t-g(\varphi),
    \end{equation}
where $\varphi\in[0,2\pi]$ and $-\infty<t<\infty$. Analytically continuing\footnote{For more details on analytic continuations in WCFTs see e.g. \cite{Detournay:2012pc,Castro:2015uaa,Song:2017czq,Jensen:2017tnb}.} $x\rightarrow -i x$ one can introduce complex coordinates $\varphi\rightarrow-z$ and $t\rightarrow\bar{z}$. One way to compute the Ward identities in the presence of sources as shown for example in \cite{Gutperle:2011kf} is to add appropriate source terms to the euclidean path integral and to compute the one point functions of the WCFT spin-2 current $\mathcal{T}(z)$ and spin-1 current $\mathcal{P}(z)$. The insertion of the additional source terms causes the WCFT currents to pick up an additional $\bar{z}$ dependence. Therefore, we are interested in computing
    \begin{equation}
        \partial_{\bar{z}}\langle\mathcal{T}(z,\bar{z})\rangle_{\mu,\nu},\qquad\partial_{\bar{z}}\langle\mathcal{P}(z,\bar{z})\rangle_{\mu,\nu},
    \end{equation}
where $\langle\ldots\rangle_{\mu,\nu}$ denotes an insertion of $e^{\frac{1}{2\pi}\int\left(\mu\mathcal{T}+\nu\mathcal{P}\right)}$ inside the expectation value. Using
    \begin{subequations}
        \begin{align}
            \mathcal{T}(z)\mathcal{T}(w) & \sim \frac{c/2}{(z-w)^4}+\frac{2\mathcal{T}(w)}{(z-w)^2}+\frac{\partial_w\mathcal{T}(w)}{z-w},\\
            \mathcal{T}(z)\mathcal{P}(w) & \sim \frac{\mathcal{P}(w)}{(z-w)^2}+\frac{\partial_w\mathcal{P}(w)}{z-w},\\
            \mathcal{P}(z)\mathcal{P}(w) & \sim \frac{\kappa/2}{(z-w)^2},
        \end{align}
    \end{subequations}
and the relation
    \begin{equation}
        \partial_{\bar{z}}\left(\frac{1}{z}\right)=2\pi\delta^{(2)}(z,\bar{z}),
    \end{equation}
as well as expanding in powers of $\mu$ and $\nu$, one obtains
    \begin{subequations}
        \begin{align}
            \partial_{\bar{z}}\langle\mathcal{T}\rangle_{\mu,\nu} & =-\langle\mu\partial_z\mathcal{T}+2\mathcal{T}\partial_z\mu-\frac{c}{12}\partial_z^3\mu+\mathcal{P}\partial_z\nu\rangle_{\mu,\nu} ,\\
            \partial_{\bar{z}}\langle\mathcal{P}\rangle_{\mu,\nu} & = -\langle\mu\partial_z\mathcal{P}+\mathcal{P}\partial_z\mu+\frac{\kappa}{2}\partial_z\nu\rangle_{\mu,\nu}.
        \end{align}
    \end{subequations}
Inverting the analytic continuation that lead to $z$ and $\bar{z}$, i.e. going back to $\varphi$ and $t$ coordinates, one obtains
    \begin{subequations}\label{eq:WardId}
        \begin{align}
            \partial_{t}\langle\mathcal{T}\rangle_{\mu,\nu} & =\langle\mu\partial_\varphi\mathcal{T}+2\mathcal{T}\partial_\varphi\mu-\frac{c}{12}\partial_\varphi^3\mu+\mathcal{P}\partial_\varphi\nu\rangle_{\mu,\nu} ,\\
            \partial_{t}\langle\mathcal{P}\rangle_{\mu,\nu} & = \langle\mu\partial_\varphi\mathcal{P}+\mathcal{P}\partial_\varphi\mu+\frac{\kappa}{2}\partial_\varphi\nu\rangle_{\mu,\nu}.
        \end{align}
    \end{subequations}
Upon identifying $\mathcal{T}=2\pi\mathcal{L}$, one obtains precisely the field equations \eqref{eq:CSEOM}. It is also straightforward to check that the Ward identities in the presence of sources \eqref{eq:WardId} reduce to the usual WCFT Ward identities $\partial_t\mathcal{T}=0$ and $\partial_t\mathcal{P}=0$ \cite{Detournay:2012pc} when setting $\mu$ and $\nu$ to zero. This shows that the Chern-Simons equations of motion -- as in the more well-known AdS$_3$ case -- indeed correctly encode the Ward identities of the dual quantum field theory.

In addition to this one still has to prove that $\mathcal{L}$ and $\mathcal{K}$ represent indeed the correct one point functions in the presence of the chemical potentials $\mu$ and $\nu$ \cite{Banados:2004nr}. For that, one has to show that for arbitrary and most importantly \emph{not} fixed i.e. $\delta\mu\neq0$ and $\delta\nu\neq0$, the variation of the total action $I_{\textrm{tot}}$ on-shell has to satisfy the schematic relation \eqref{eq:SchematicOnShellVariation}. For the boundary conditions \eqref{eq:WAdSBCs} one obtains
    \begin{equation}
        \delta I_\textrm{B}=-\int_{\partial\mathcal{M}}\left(\mathcal{L}\delta\mu+\mathcal{K}\delta\nu\right).
    \end{equation}
and thus one can identify $\mathcal{L}$ and $\mathcal{K}$ with the vacuum expectation values of the WCFT energy-momentum tensor and $\hat{\mathfrak{u}}(1)$ current and $\mu$ and $\nu$ as the corresponding sources.

\subsection{Thermal entropy: Mass, angular momentum and holonomies}\label{sec:EntropyMassAngularMomentumAndHolonmies}

We now proceed in showing that the boundary conditions \eqref{eq:WAdSBCs} do contain spacelike warped AdS$_3$ black holes. In this section we do this by determining the thermal entropy of the configuration \eqref{eq:WAdSBCs}. We note that from now on we 
will assume that the chemical potentials $\mu$ and $\nu$  as well as
the state-dependent functions $\mathcal{L}$ and $\mathcal{K}$ are constant. Under this assumption, the connections $a_t$ and $c_t$ simplify as
\begin{equation}
		a_t=\mu \Lt_1-\mathfrak{L}\mu\Lt_{-1},\qquad c_t=\left(\nu+\frac{4\pi}{\kappa}\mathcal{K}\mu\right)\St.
	\end{equation}
Starting with this setup, the procedure in determining the thermal entropy of such solutions is roughly as follows:
	\begin{itemize}
		\item Identify mass and angular momentum with the charges that generate time and angular translations respectively.
		\item Impose suitable holonomy conditions to fix the inverse temperature and angular velocity as functions of mass, angular momentum.
		\item Integrate the first law of black hole thermodynamics to obtain the thermal entropy.
	\end{itemize}
Some comments are in order. The first comment is related to determining the mass and angular momentum of the configuration \eqref{eq:WAdSBCs}. In the usual metric formulation of Einstein gravity, the mass and angular momentum are associated with the charges of the Killing vectors $\partial_t$ and $\partial_\varphi$, respectively, and there is a precise way of relating these charges with the ones determined in the Chern-Simons formulation (see e.g. \cite{Henneaux:1984ji,Bunster:2014cna,Perez:2015jxn}). This, however, requires some geometric input to be sure that one is identifying the correct quantities as mass and angular momentum in the Chern-Simons formulation. In this section we will require that a similar relation holds also in the case at hand. We will show later in Sec.~\ref{Sec:Metric} that this requirement is indeed valid.

The second comment is related to the holonomy conditions we are proposing below in this section. At first we will motivate these conditions based on the first law of of black hole thermodynamics. We stress that the exact form of these conditions at this point of the computation is, similar to the way we determine mass and angular momentum, an educated guess in the absence of a geometric interpretation. In Sec.~\ref{Sec:Metric} we will argue that these holonomy conditions are, indeed, a sensible choice by determining the inverse temperature and angular velocity using a metric interpretation of the boundary conditions \eqref{eq:WAdSBCs} and showing that these expression match precisely with the ones obtained from the Chern-Simons description with the proposed holonomy conditions. In addition, one can also use WCFT arguments to show that the proposed holonomy conditions yield the expressions for inverse temperature and angular velocity expected from a WCFT perspective.

Now we move to determining the mass and angular momentum. In a Chern-Simons description of three-dimensional spacetimes such as AdS$_3$, the gauge parameters $\varepsilon$ preserving the connection $\mathcal{A}$ are related on-shell to the Killing vectors $\xi^\mu$ of the corresponding spacetime via the relation $\epsilon=\xi^\mu\mathcal{A}_\mu$. Since the description used in this work is similar to the situations encountered in e.g. \cite{Henneaux:1984ji,Bunster:2014cna,Perez:2015jxn}, it is reasonable to that the gauge parameters \eqref{eq:GaugeParameter} preserving the connection \eqref{eq:WAdSBCs} are related to the corresponding Killing vectors via $\varepsilon=\xi^\mu\mathcal{A}_\mu$, $\bar{\varepsilon}=\xi^\mu\mathcal{C}_\mu$. With this requirement one can determine (the variation of) mass and angular momentum of the solutions \eqref{eq:WAdSBCs} via
	\begin{subequations}\label{eq:VariationMassAngMomFormula}
    	\begin{align}
        		\delta M &:= \delta Q[\varepsilon\big|_{\partial_t}]+\delta Q[\bar{\varepsilon}\big|_{\partial_t}]
                =\frac{k}{2\pi}\int\extd\varphi\langle\mathcal{A}_t\delta\mathcal{A}_\varphi\rangle+\frac{\kappa}{4\pi}\int\extd\varphi\langle\mathcal{C}_t\delta\mathcal{C}_\varphi\rangle,\\
        		\delta J &:= \delta Q[\varepsilon\big|_{-\partial_\varphi}]+\delta Q[\bar{\varepsilon}\big|_{-\partial_\varphi}]=-\frac{k}{2\pi}\int\extd\varphi\langle\mathcal{A}_\varphi\delta\mathcal{A}_\varphi\rangle-\frac{\kappa}{4\pi}\int\extd\varphi\langle\mathcal{C}_\varphi\delta\mathcal{C}_\varphi\rangle.
    	\end{align}
    	\end{subequations}
For the boundary conditions \eqref{eq:WAdSBCs} one obtains the following expressions
 	\begin{equation}
 		\delta M = 2\pi\left(\mu\,\delta\mathcal{L}+\nu\delta\mathcal{K}\right),\qquad
 		\delta J = -2\pi \delta\mathcal{L}.
 	\end{equation}
Fixing the chemical potentials $\mu$ and $\nu$ is tantamount to fixing ``units'' to measure the energy. One possible choice for example is $\mu=0$ and $\nu=1$ for which mass and angular momentum 
are given by 
	\begin{equation}\label{eq:MassAndAngularMomentum}
		M=2\pi\mathcal{K},\qquad J=-2\pi \mathcal{L}.
	\end{equation}
For this choice of $\mu$ and $\nu$, one exactly recovers the boundary conditions of \cite{Hofman:2014loa}. Thus in this case the boundary conditions found in \cite{Hofman:2014loa} also contain spacelike warped black hole solutions. We would also like to stress that the existence of these black hole solutions was not realized in  \cite{Hofman:2014loa}.

If the connection \eqref{eq:WAdSBCs} describes a warped AdS$_3$ black hole, then the thermal entropy also has to satisfy the first law of black hole thermodynamics
	\begin{equation}\label{eq:FirstLaw}
		\delta S_{\textrm{Th}}=\beta\left(\delta M-\Omega\delta J\right),
	\end{equation}
where $S_{\textrm{Th}}$ is the thermal entropy, $\beta$ the inverse temperature and $\Omega$ the angular velocity. We have already determined what (the variation of) mass and angular momentum is in the previous paragraph. The only missing ingredients in order to determine the variation of the thermal entropy $\delta S_{\textrm{Th}}$ are the functional relations between the inverse temperature $\beta$, the angular velocity $\Omega$ and the mass $M$ and angular momentum $J$. Once these relations are identified, one can functionally integrate \eqref{eq:FirstLaw} to obtain the thermal entropy.

In a Chern-Simons theory these functional relations are usually determined by looking at the holonomy of the connection (see e.g. \cite{Gutperle:2011kf,Ammon:2011nk,deBoer:2013gz,Bunster:2014mua}). Thus, in what follows we will be choosing certain holonomy conditions that will fix $\beta$ and $\Omega$ in terms of the state-dependent functions $\mathcal{L}$ and $\mathcal{K}$. Before stating those conditions, it will be illuminating to rewrite the first law \eqref{eq:FirstLaw} as
    \begin{align}\label{eq:VariationEntropy}
        \delta S_{\textrm{Th}}=&\frac{k}{2\pi}\beta\int\extd\varphi \langle a_t\delta a_\varphi\rangle+\frac{k}{2\pi}\beta\,\Omega\int\extd\varphi \langle a_\varphi\delta a_\varphi\rangle\nonumber\\
        &+\frac{\kappa}{4\pi}\beta\int\extd\varphi \langle c_t\delta c_\varphi\rangle+\frac{\kappa}{4\pi}\beta\,\Omega\int\extd\varphi \langle c_\varphi\delta c_\varphi\rangle,
    \end{align}    
or, in a little bit more suggestive manner as
    \begin{equation}\label{eq:EntropyAndHol2}
        \delta S_{\textrm{Th}} = k\langle h\,\delta a_\varphi\rangle+\frac{\kappa}{2}\langle \bar{h}\,\delta 
        c_\varphi\rangle,
    \end{equation}
with 
    \begin{subequations}\label{eq:HolonomyDefinition}
    \begin{align}
        h & =\frac{\beta}{2\pi}\left(\int\extd\varphi\,a_t + \Omega \int\extd\varphi\,a_\varphi\right),\label{eq:HolonomyDefinition1}\\
        \bar{h} & =\frac{\beta}{2\pi}\left(\int\extd\varphi\,c_t + \Omega \int\extd\varphi\,c_\varphi\right).\label{eq:HolonomyDefinition2}
    \end{align}
    \end{subequations} 
One can see already at this stage that if $h$ is proportional to the elements of the center of the relevant gauge algebra then two things happen:
	\begin{itemize}
		\item The expression for the variation of the entropy \eqref{eq:EntropyAndHol2} can be trivially integrated.
		\item One makes manifest that all the relevant information regarding entropy is encoded in the connections along the noncontractible cycle that wraps around the horizon. Or, in other words, the thermal entropy can be computed by using a Wilson line wrapping around the horizon \cite{Ammon:2013hba,deBoer:2013vca}.
	\end{itemize}
Moreover, the holonomies of the rotating BTZ black hole for example also take the form \eqref{eq:HolonomyDefinition1} (see \eqref{eq:BTZHolonomiesLorentzian} in Appendix~\ref{sec:AppendixA}). Thus a suggestive choice of holonomy conditions in the current case is
	\begin{equation}\label{eq:HolonomyConditions}
		\textrm{Eigen}\left[h\right]= \textrm{Eigen}\left[2\pi \Lt_0\right],\qquad\textrm{Eigen}\left[\bar{h}\right]=\textrm{Eigen}\left[2\pi \gamma \St\right],
	\end{equation}
where $\textrm{Eigen}[...]$ denotes a set of eigenvalues for $...$ and $\gamma$ is some constant. Imposing these conditions, one finds that the holonomies of $h$ and $\bar{h}$ are given by
    \begin{equation}
        b^{-1}e^{i h}b=-\unity,\quad e^{i\bar{h}}=e^{2\pi i \gamma}.
    \end{equation}
Therefore, if $\gamma$ is an integer for example one finds that the holonomy lies in the center of the gauge group, in close analogy to the BTZ case. However, since our goal is to describe a warped geometry, we do not assume that this has to be necessarily the case. The reason is that warped AdS$_3$ spacetimes are deformations of AdS$_3$ spacetimes and, as such, it is suggestive that a similar kind of ``deformation'' also happens at the level of holonomies.

Enforcing the conditions \eqref{eq:HolonomyConditions} fixes the inverse temperature and angular velocity as
	\begin{equation}\label{eq:InvTempAndAngPotPrelim}
		\beta =\frac{2\pi}{\nu}\left(\gamma-\frac{2\pi\mathcal{K}}{\kappa\sqrt{\mathfrak{L}}}\right),\qquad
		\Omega =\frac{\nu}{2\gamma\sqrt{\mathfrak{L}}-\frac{4\pi\mathcal{K}}{\kappa}}-\mu.
	\end{equation}
Both the inverse temperature and the angular velocity are theory-independent quantities and only depend on the geometry in question. Thus, we will use the metric interpretation of Sec.~\ref{Sec:Metric} later on to fix the exact value of $\gamma$.

After this, one can directly proceed in determining the thermal entropy via functionally integrating \eqref{eq:EntropyAndHol2} to obtain
	\begin{equation}\label{eq:CSEntropyOld}
		S_{\rm Th}=2\pi\left(2\pi\mathcal{K}\gamma+\sqrt{\frac{c}{6}\left(2\pi\mathcal{L}-\frac{4\pi^2\mathcal{K}^2}{\kappa}\right)}\right).
	\end{equation}
For the choice $\mu=0$ and $\nu=1$, the thermodynamic potentials are given by
	\begin{equation}\label{eq:InvTempAndAngPot}
		\beta =2\pi\left(\gamma-\frac{2\pi\mathcal{K}}{\kappa\sqrt{\mathfrak{L}}}\right),\qquad
		\Omega =\frac{1}{2\gamma\sqrt{\mathfrak{L}}-\frac{4\pi\mathcal{K}}{\kappa}},
	\end{equation}
and one obtains the following thermal entropy in terms of mass and angular momentum:
	\begin{equation}\label{eq:CSEntropy}
		S_{\rm Th}=2\pi\left(M\gamma+\sqrt{\frac{c}{6}\left(-J-\frac{M^2}{\kappa}\right)}\right),
	\end{equation}
where $c=6k$. This is exactly the form of the entropy for a spacelike warped AdS$_3$ black hole and at the same time that of a WCFT at finite temperature \cite{Detournay:2012pc}. Since for this choice of chemical potentials the mass and angular momentum are directly related to the zero modes of the functions $\mathcal{L}$ and $\mathcal{K}$, we will assume that $\mu=0$ and $\nu=1$ in everything that will follow from now on. This expression for the entropy is already a strong evidence supporting that the connection \eqref{eq:WAdSBCs} correctly describes spacelike warped AdS$_3$ black holes.

In the rest of this paper, we will show that the holonomy conditions \eqref{eq:HolonomyConditions} are sensible choices for spacelike warped AdS$_3$ black holes.

\section{Vacuum state and WCFT entropy}\label{Sec:VacuumCSWCFT}

The expression for the entropy \eqref{eq:CSEntropy} takes the form expected for the thermal entropy of a WCFT. A natural question is then how to relate these two quantities to each other. The general formula for the thermal entropy for a WCFT at finite temperature is given by \cite{Detournay:2012pc}
    \begin{equation}\label{eq:WCFTEntropy}
        S_{\textrm{Th}} = -\frac{4\pi i M M^{\textrm v}}{\kappa}+4\pi\sqrt{\left(J^{\textrm v}+\frac{\left(M^{\textrm v}\right)^2}{\kappa}\right)\left(J+\frac{M^2}{\kappa}\right)}.
    \end{equation}
In this expression, $M$ and $J$ are mass and angular momentum, respectively and $M^v$ and $J^v$ are those of the vacuum. Thus, in order to make contact with this formula, one first needs to determine what the vacuum solution of \eqref{eq:CSAction} is.

In the usual metric formulation of Einstein gravity, one way to determine the vacuum is via the maximal number of globally well defined Killing vectors. Looking at the BTZ black hole for example one finds in general six linearly-independent Killing vectors.\footnote{For more details please take a look at Appendix~\ref{sec:AppendixA}, specifically \eqref{eq:BTZKillingVectors}.} However, out of these six Killing vectors, only two are globally well defined for general values of mass and angular momentum: the ones associated to time and angular translations. One possible definition of a vacuum state is that it is the state with the highest amount of symmetry. Thus, in this case it should be the state where all the six Killing vectors are well defined globally. For the BTZ black hole, this happens for very specific values of the mass and angular momentum (see \eqref{eq:BTZHolonomiesVacuum}) and yields global AdS$_3$.

For spacelike warped AdS$_3$ black holes, the situation is similar; the only difference is that one generically has four Killing vectors out of which two are again globally well defined for any value of mass and angular momentum. In a Chern-Simons theory the role of Killing vectors is taken by the gauge parameters \eqref{eq:WAdSGaugeTrafos} and as such one might expect that one encounters similar features at the level of the gauge parameters. Another way to look for the vacuum in a Chern-Simons formulation is to use the holonomies. Taking again the BTZ black hole as inspiration (see \eqref{eq:BTZHolonomiesVacuumConditions}), it is straightforward to check that the holonomies around the $\varphi$-cycle are in general nontrivial. Only for very specific values of mass and angular momentum -- the ones that give global AdS$_3$ -- these holonomies become trivial. In the following we will use this as a guiding principle to determine the vacuum state of the solution \eqref{eq:WAdSBCs}.

Inspired by this, we require that, for the vacuum state of the warped AdS$_3$ case
the holonomy along the $\varphi$-cycle obeys
    \begin{equation}\label{eq:CSVacuumHolonomy}
        b^{-1}e^{\oint a_\varphi}b=-\unity,\quad e^{\oint c_\varphi}=e^{2\pi i \gamma}.
    \end{equation}
This leads to the following restrictions of the state-dependent functions,
    \begin{equation}
        \mathcal{L}^{\textrm v}=-\frac{c(1+2n)^2}{48\pi}-\frac{\gamma^2\kappa}{8\pi},\quad \mathcal{K}^{\textrm v}= \frac{i\kappa \gamma}{4\pi},
    \end{equation}
or on terms of mass and angular momentum (again for the choice $\mu=0$, $\nu=1$)
    \begin{equation}\label{eq:CSVacuum}
        J^{\textrm v}=\frac{c(1+2n)^2}{24}+\frac{\gamma^2\kappa}{4},\quad M^{\textrm v}= \frac{i\kappa \gamma}{2}.
    \end{equation}
The integer $n$ accounts for the periodicity of the complex exponential function and leads to additional possible branches of solutions.\footnote{This also happens already in the BTZ case, see \eqref{eq:BTZHolonomiesVacuum}.} In the following we will choose the branch with $n=0$ in order to make contact with the WCFT vacuum values found in \cite{Detournay:2012pc}. We note that this is a natural choice which avoids a multiple cover around the $\varphi$-cycle.

In order to check if the vacuum values lead to some kind of symmetry enhancement at the level of the gauge parameters,  we first determine these gauge parameters satisfying $\delta_\varepsilon\mathcal{A}=\delta_{\bar{\varepsilon}}\mathcal{C}=0$. From these equations, one finds that the gauge generators defined in \eqref{eq:GaugeParameter} satisfy
    \begin{equation}
        \epsilon'''  =\frac{8\pi}{k}\left(\mathcal{L}-\frac{2\pi}{\kappa}\mathcal{K}^2\right)\epsilon',\qquad \partial_t\epsilon  =\mu\epsilon',\qquad
        \partial_t\sigma =-\frac{4\pi}{\kappa}\mu\mathcal{K}\epsilon'=\mu\sigma'.
    \end{equation}
For $\mu=0$ and $\nu=1$, the solutions to these differential equations are given by
    \begin{equation}\label{eq:VacuumGaugeParameters}
        \epsilon =\frac{e^{2\sqrt{\mathfrak{L}}\varphi}C_1-e^{-2\sqrt{\mathfrak{L}}\varphi}C_2}{2\sqrt{\mathfrak{L}}}+C_3,\qquad
        \sigma = -\frac{4\pi}{\kappa}\mathcal{K}\epsilon+C_4.
    \end{equation}
Since the manifold $\mathcal{M}$ has the topology of a solid cylinder whose $\varphi$ coordinate is $2\pi$-periodic, one also has to require that $\epsilon(\varphi)\sim\epsilon(\varphi+2\pi)$ as well as $\sigma(\varphi)\sim\sigma(\varphi+2\pi)$. However, looking at \eqref{eq:VacuumGaugeParameters} one sees that for generic values of $\mathfrak{L}$ this is only true if $C_1=C_2=0$. This fits perfectly into the picture of a spacelike warped black hole: out of the four linearly independent Killing vectors, only two are globally well defined. On the other hand, if $\mathcal{L}$ and $\mathcal{K}$ take the values found in \eqref{eq:CSVacuum}, one sees that $\sqrt{\mathfrak{L}}=\frac{i}{2}$ and, thus,
    \begin{equation}\label{eq:VacuumGaugeParametersInsert}
        \epsilon =-i\left(e^{i\varphi}C_1-e^{-i\varphi}C_2\right)+C_3,\qquad
        \sigma = -i\gamma\epsilon+C_4.
    \end{equation}
These gauge parameters are $2\pi$-periodic for any values of the four integration constants $C_n$. This fits perfectly with what we expected at the beginning of this section: there should be four linearly independent globally well defined gauge parameters for the vacuum state.  

After having determined the vacuum, one can now make contact with the general formula of a WCFT at finite temperature \eqref{eq:WCFTEntropy}. Plugging the vacuum values \eqref{eq:CSVacuum} into that expression one obtains
	\begin{equation}
		S_{\rm Th}=2\pi\left(M\gamma+\sqrt{\frac{c}{6}\left(-J-\frac{M^2}{\kappa}\right)}\right).
	\end{equation}
This is exactly the same form as the entropy derived in the gravitational setup \eqref{eq:CSEntropy}. We note that \eqref{eq:CSVacuum} indicates the parameter $\gamma$ introduced in \eqref{eq:HolonomyConditions} is related to the vacuum value of the mass.

Having shown that the vacuum values \eqref{eq:CSVacuum} in combination with the general WCFT formula \eqref{eq:WCFTEntropy} yields the same expression as the thermal entropy \eqref{eq:CSEntropy}, we will now present a WCFT argument that supports the choice of holonomy conditions \eqref{eq:HolonomyConditions}. Starting from the WCFT formula \eqref{eq:WCFTEntropy} one can determine its variation with respect to the charges $J$ and $M$. This allows one to identify the inverse temperature $\beta$ and angular velocity $\Omega$ completely in terms of vacuum values and charges by using the first law \eqref{eq:FirstLaw}. This yields the following expressions:
    \begin{subequations}
    \begin{align}
        \beta = & -\frac{4\pi i}{\kappa}M^{\textrm v}+\frac{4\pi\left(-J^{\textrm v}-\frac{\left(M^{\textrm v}\right)^2}{\kappa}\right)M}{\kappa\sqrt{-\left(-J^{\textrm v}-\frac{\left(M^{\textrm v}\right)^2}{\kappa}\right)\left(-J-\frac{M^2}{\kappa}\right)}},\\
        \beta\,\Omega = &- \frac{2\pi\left(-J^{\textrm v}-\frac{\left(M^{\textrm v}\right)^2}{\kappa}\right)}{\sqrt{-\left(-J^{\textrm v}-\frac{\left(M^{\textrm v}\right)^2}{\kappa}\right)\left(-J-\frac{M^2}{\kappa}\right)}}.
    \end{align}    
    \end{subequations}
Inserting the vacuum values \eqref{eq:CSVacuum} one immediately recovers \eqref{eq:InvTempAndAngPot}. This is additional evidence supporting the validity of \eqref{eq:HolonomyConditions}.
\section{Holographic entanglement entropy and thermal entropy from wilson lines}\label{Sec:HEEWilson}

In this section we compute the holographic entanglement entropy of our setup by using Wilson lines\footnote{Wilson lines are in general very versatile objects to consider in gauge theories. Another interesting application can for example be found in \cite{Castro:2016ehj} where the authors used Wilson lines to probe Lorentzian eternal higher-spin black holes in AdS$_3$.} as described in \cite{Castro:2015csg}. With the Wilson line method, we will also give an alternative way of computing the thermal entropy of the configuration \eqref{eq:WAdSBCs}. As we will see, this computation also helps us to understand the physical meaning 
of the parameter $\gamma$. In addition this computation gives an additional nontrivial check that the holonomy conditions \eqref{eq:HolonomyConditions} are, indeed, a sensible choice.

\subsection{Holographic entanglement entropy}

The concept that entanglement entropy can be holographically computed by using extremal surfaces in the bulk has been first made precise in \cite{Ryu:2006bv,Ryu:2006ef} and then subsequently generalized in \cite{Hubeny:2007xt}. The fact that one can compute entanglement entropy holographically gives a very beautiful and intriguing relation between geometry and quantum information. In order to extend these original ideas of holographic entanglement entropy also to higher-spin theories in three dimensions it turns out that the natural generalization in the context of Chern-Simons theories is to use Wilson lines. For Chern-Simons theories with $\mathfrak{sl}(N,\mathbb{R})\oplus\mathfrak{sl}(N,\mathbb{R})$ symmetries, the basic idea to use Wilson lines 
for computing holographic entanglement entropy has been first made precise in \cite{Ammon:2013hba,deBoer:2013vca}\footnote{For the case of $\mathfrak{isl}(N,\mathbb{R})$ that is relevant for flat space holography see \cite{Bagchi:2014iea,Basu:2015evh}. For the dual CFT computation including spin-3 currents that put the proposal of \cite{Ammon:2013hba,deBoer:2013vca} on an even more solid footing see e.g. \cite{Datta:2014ska,Datta:2014uxa}}. 
For 
$\mathfrak{sl}(2,\mathbb{R})\oplus\mathfrak{u}(1)$ Chern-Simons theories that should be dual to WCFTs, 
slight modifications of the methods by \cite{Ammon:2013hba,deBoer:2013vca} are necessary and have been done in \cite{Castro:2015csg}. Since one can find the detailed construction already very well explained in \cite{Castro:2015csg}, we will in the following only review the main ideas of the construction and focus on the parts that are essential for the purpose of our work.

\begin{figure}[ht]
		\centering
		\includegraphics[width=0.5\textwidth]{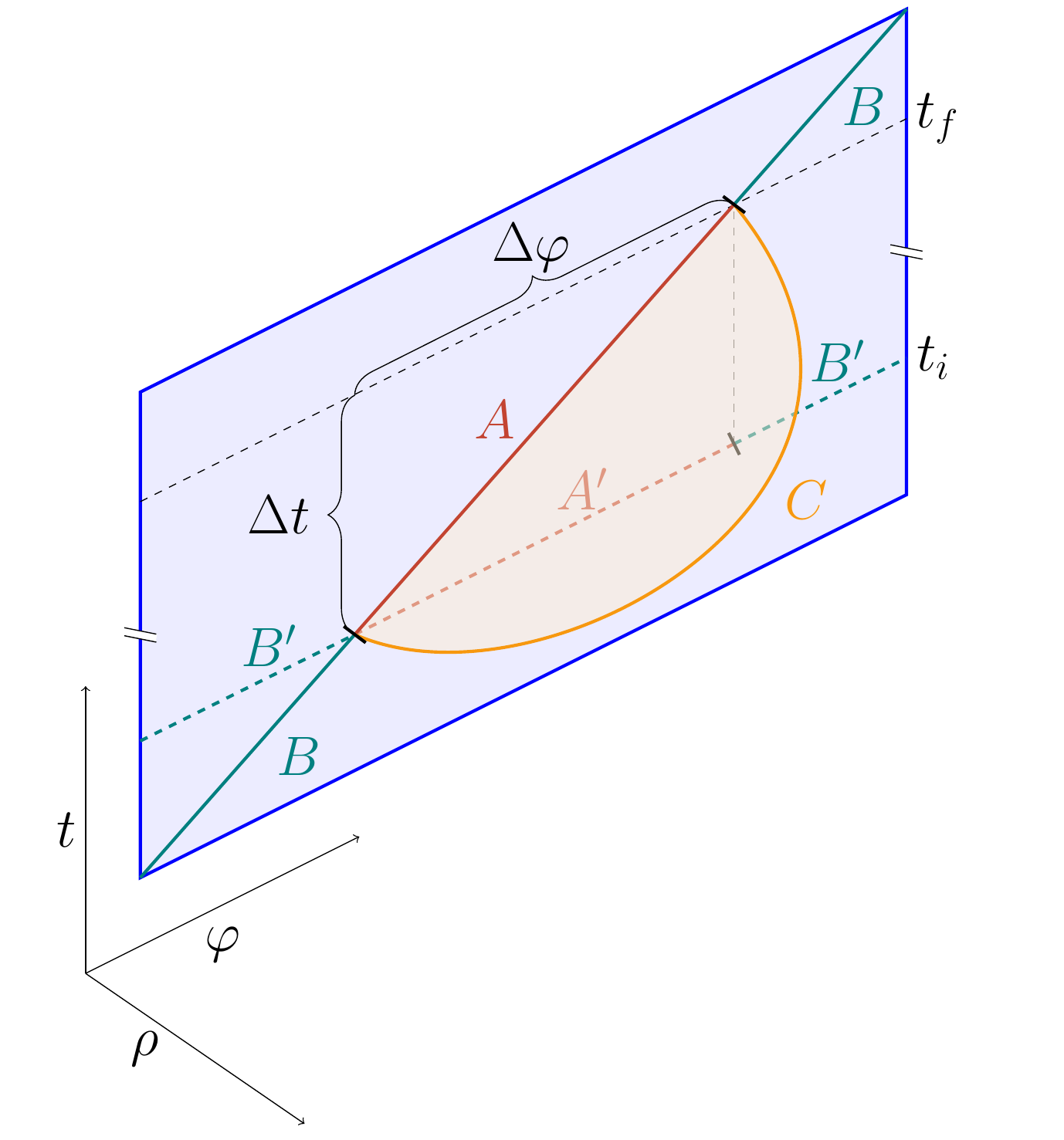}
		\caption{Boosted ({\color{BrickRed!80}$A$}, {\color{teal}$B$}) and equal time ({\color{BrickRed!80}$A'$}, {\color{teal}$B'$}) entangled intervals and the corresponding Wilson line ({\color{YellowOrange}$C$}) used to determine holographic entanglement entropy for WCFTs.}
		\label{fig:HEEPicture}
	\end{figure}

The main idea of computing holographic entanglement entropy in WCFTs is using Wilson lines. To be more precise, the negative logarithm of the trace of a Wilson line taken in an appropriate representation computes the entanglement entropy of the boundary region bounded by the endpoints of the Wilson line. In the case of WCFTs, this means 
    \begin{equation}
        S_{\textrm{EE}}=-\log \left[\mathcal{W}_\mathcal{R}^{\mathfrak{sl}(2,\mathbb{R})}(C;\mathcal{A})\right]-\log \left[\mathcal{W}_\mathcal{R}^{\mathfrak{u}(1)}(C;\mathcal{C})\right],
    \end{equation}
where $C$ denotes the path of the Wilson line and $\mathcal{R}$ is the chosen representation (usually an infinite-dimensional representation). A more physical interpretation of this procedure is based on a massive (and spinning) particle travelling along the bulk: the path $C$ describes 
the world line of the massive particle. For WCFTs, it was argued in \cite{Castro:2015csg} that the appropriate representations for the $\mathfrak{sl}(2,\mathbb{R})$ part are given by a single particle living on AdS$_2$. Following the construction in \cite{Castro:2015csg}, one can find that the leading order piece for the entanglement entropy of the $\mathfrak{sl}(2,\mathbb{R})$ part of the Wilson line that is attached to some cutoff surface $\rho_0$ very close to the boundary is essentially given by
    \begin{equation}\label{eq:SL2HEEFormula}
        S_{\textrm{EE}}^{\mathfrak{sl}(2,\mathbb{R})}=-2c_2\Delta\alpha.
    \end{equation}
Here $c_2$ is the quadratic Casimir 
of $\mathfrak{sl}(2,\mathbb{R})$ and $\Delta\alpha$ is determined via
    \begin{equation}
        2\cosh\left[\Delta\alpha\sqrt{2c_2}\right]=2\sqrt{1-\textrm{tr}(\mathfrak{M}\,\Lt_+)\textrm{tr}(\mathfrak{M}\,\Lt_-)},
    \end{equation}
with
    \begin{equation}\label{eq:EntanglementMatrix}
        \mathfrak{M}=g_f^{-1}L(\rho_0,\varphi_f,t_f)L^{-1}(\rho_0,\varphi_i,t_i)g_i,
    \end{equation}
where
    \begin{subequations}
    \begin{align}
        g_f & =g_i=\exp\left[-\frac{\pi}{4}\left(\Lt_+-\Lt_-\right)\right],\\ 
        L(\rho,\varphi,t) & =\exp\left[-\rho\Lt_0\right]\exp\left[-a_\varphi\varphi-a_tt\right].
    \end{align}
    \end{subequations}
At this point it is important to note that a WCFT is not a relativistic quantum field theory. As such entanglement entropy will look different for different observers. Thus, the Wilson line is attached at the initial point ($\rho_0,\varphi_i,t_i$) and the final point ($\rho_0,\varphi_f,t_f$). See Fig.~\ref{fig:HEEPicture} for a graphic depiction.     
For the connection \eqref{eq:WAdSBCs}, under the assumption that the entangling interval $\varphi_f-\varphi_i\equiv\Delta\varphi$ is very large compared to the UV cutoff $\epsilon=2e^{-\rho_0}$ i.e. $\frac{\Delta\varphi}{\epsilon}\gg1$,  one finds that
    \begin{equation}
        2\cosh\left[\Delta\alpha\sqrt{2c_2}\right]=2\frac{\beta_\varphi}{\pi\epsilon}\sinh\left[\frac{\pi\Delta\varphi}{\beta_\varphi}\right],
    \end{equation}
where
    \begin{equation}\label{eq:BetaPhi}
        \beta_\varphi=\frac{\pi}{\sqrt{\mathfrak{L}}}=\frac{\pi}{\sqrt{\frac{6}{c}\left(-J-\frac{M^2}{\kappa}\right)}},
    \end{equation}
and $c=6k$. Identifying the quadratic Casimir in terms of the central charge as $c_2=\frac{k^2}{2}=\frac{c^2}{72}$ and taking the semiclassical limit i.e. $k\rightarrow\infty$, one can solve for $\Delta\alpha$ and, thus, obtain the following expression for the $\mathfrak{sl}(2,\mathbb{R})$ part of the entanglement entropy\footnote{Since the $\cosh$ is an even function one obtains different signs depending on the sign of $\Delta\alpha$. Here we chose to solve for the branch where the entanglement entropy has a positive overall sign.}
    \begin{equation}
        S_{\textrm{EE}}^{\mathfrak{sl}(2,\mathbb{R})}=\frac{c}{6}\log\left[\frac{\beta_\varphi}{\pi\epsilon}\sinh\left[\frac{\pi\Delta\varphi}{\beta_\varphi}\right]\right].
    \end{equation}
The $\mathfrak{u}(1)$ part of the computation is simpler to perform since the Wilson line in this case is given by the integral along the world line of the massive particle as\footnote{One might wonder why the prefactor in front of the integral is proportional to the vacuum value of the mass. One way to see this is by arguing that the resulting expression for the entanglement entropy should be in accordance with the general formula \eqref{eq:HEESong}.}
    \begin{equation}
        \mathcal{W}_\mathcal{R}^{\mathfrak{u}(1)}(C;\mathcal{C})=\exp\left[iM^{\textrm{v}}\int_C\mathcal{C}\right].
    \end{equation}
Thus one obtains as the full expression for the holographic  entanglement entropy
    \begin{equation}\label{eq:HEEFull1}
        S_{\textrm{EE}}=-iM^{\textrm{v}}\left(\Delta t+2\frac{M}{\kappa}\Delta\varphi\right)+\frac{c}{6}\log\left[\frac{\beta_\varphi}{\pi\epsilon}\sinh\left[\frac{\pi\Delta\varphi}{\beta_\varphi}\right]\right]. 
    \end{equation}
After using the vacuum expressions for $M^v$ found in \eqref{eq:CSVacuum}, 
this turns into
    \begin{equation}\label{eq:HEEFull2}
        S_{\textrm{EE}}=\gamma\left(\frac{\kappa}{2}\Delta t+M\Delta\varphi\right)+\frac{c}{6}\log\left[\frac{\beta_\varphi}{\pi\epsilon}\sinh\left[\frac{\pi\Delta\varphi}{\beta_\varphi}\right]\right].
    \end{equation}
It is instructive to compare the expression \eqref{eq:HEEFull1} with the general expression\footnote{Please note that there is a sign difference with respect to the time interval $\Delta T$ used in \cite{Song:2016gtd} and the time interval $\Delta t$ used here. The conventions for time used in this work are consistent with the ones used in \cite{Castro:2015csg}.} derived in \cite{Song:2016gtd}
    \begin{equation}\label{eq:HEESong}
        S_{\textrm{EE}}=iM^{\textrm{v}}\left(-\Delta t+\frac{\beta-\delta}{\beta_\varphi}\Delta\varphi\right)+\left(i\frac{\delta}{\pi}M^{\textrm{v}}-4J^{\textrm{v}}\right)\log\left[\frac{\beta_\varphi}{\pi\epsilon}\sinh\left[\frac{\pi\Delta\varphi}{\beta_\varphi}\right]\right].
    \end{equation}
In this formula $\delta$ is a parameter related to the tilt of the cylinder on which the dual WCFT is defined. One can reproduce our expression \eqref{eq:HEEFull1} for entanglement entropy by substituting
$\beta$ \eqref{eq:InvTempAndAngPotPrelim} and $\beta_\varphi$ \eqref{eq:BetaPhi} into \eqref{eq:HEESong} and then
using the identification $\delta=2\pi \gamma$. Thus the (seemingly) arbitrary parameter $\gamma$ encountered before is nothing but the tilting parameter of the cylinder. This is also in good agreement with the vacuum values \eqref{eq:CSVacuum}, since the origin of the nonzero vacuum value of the mass is precisely the tilt of the WCFT cylinder.

The holographic entanglement entropy of our configuration \eqref{eq:WAdSBCs} provides us with additional evidence for the validity of the holonomy conditions \eqref{eq:HolonomyConditions}. To see this, we notice that the thermal entropy \eqref{eq:CSEntropy} can also be recovered from the entanglement entropy at a given constant time slice in the limit where the entanglement entropy becomes extensive (see e.g. \cite{deBoer:2013vca}). That is, in the limit where $\frac{\Delta\varphi}{\beta_\varphi}\gg1$ one has
    \begin{equation}
        S_{\textrm{EE}}\approx\frac{S_{\textrm{Th}}}{2\pi}\Delta\varphi.
    \end{equation}
Thus, one has a way of determining the thermal entropy without any reference to the chosen holonomy conditions. Performing this extensive limit one obtains
    \begin{equation}
         S_{\textrm{EE}}\approx\left(M\gamma+\sqrt{\frac{c}{6}\left(-J-\frac{M^2}{\kappa}\right)}\right)\Delta\varphi.
    \end{equation}
Thus, the extensive limit of the entanglement entropy \eqref{eq:HEEFull2} yields again exactly the thermal entropy found previously \eqref{eq:CSEntropy}, 
justifying the validity of the holonomy conditions imposed for the 
derivation of the thermal entropy in
Sec.~\ref{sec:EntropyMassAngularMomentumAndHolonmies}.

The extensive limit is not the only interesting limit that can be taken from the entanglement entropy of our configuration. Another interesting limit is the one of small intervals i.e. $\frac{\Delta\varphi}{\beta_\varphi}\ll1$. In this limit, the entanglement entropy \eqref{eq:HEEFull2} simplifies to
    \begin{equation}\label{eq:HEESmallInt}
        S_{\textrm{EE}}\approx\gamma\frac{\kappa}{2}\Delta t+\frac{c}{6}\log\left[\frac{\Delta\varphi}{\epsilon}\right],
    \end{equation}
and becomes completely temperature independent. Precisely the same thing also happens when taking the small interval limit of entanglement entropy in an ordinary CFT at finite temperature. Thus, one would expect that also for a WCFT the vanishing temperature limit of \eqref{eq:HEESong} reproduces \eqref{eq:HEESmallInt}. And indeed, a straightforward calculation shows that this is also the case for \eqref{eq:HEESong}.
\subsection{Thermal entropy using Wilson lines}\label{sec:ThermalEntropyWilson}

One of advantages of using Wilson lines to compute entanglement entropy holographically is that one can also directly compute the thermal entropy of black hole solutions (aside using the extensive limit) by simply wrapping the Wilson line around the black hole horizon.
In the case at hand this is the $\varphi$-cycle. Thus, the initial and final points that appear in the matrix $\mathcal{M}$ in \eqref{eq:EntanglementMatrix} are now identical, and one has to solve
    \begin{equation}
        2\cosh\left[\Delta\alpha\sqrt{2c_2}\right]=2\cosh\left[\frac{2\pi^2}{\beta_\varphi}\right].
    \end{equation}
Solving for $\Delta\alpha$ and plugging the result into \eqref{eq:SL2HEEFormula}, one obtains for the $\mathfrak{sl}(2,\mathbb{R})$ part
    \begin{equation}
        S_{\textrm{Th}}^{\mathfrak{sl}(2,\mathbb{R})}=2\pi\sqrt{\frac{c}{6}\left(-J-\frac{M^2}{\kappa}\right)}.
    \end{equation}
The $\mathfrak{u}(1)$ part is again simply the world line of the massive particle around the $\varphi$-cycle and, thus, yields
    \begin{equation}
    S_{\textrm{Th}}^{\mathfrak{u}(1)}=2\pi M\gamma. 
    \end{equation}
These expressions reproduce exactly the thermal entropy \eqref{eq:CSEntropy}.

As an addendum it is worthwhile to mention that this prescription of the Wilson line wrapping the horizon is nothing else than the statement that all the relevant information about the thermal entropy is encoded in the connection  along the noncontractible cycle ($\varphi$-cycle in the current case). A convenient shortcut\footnote{See e.g. \cite{deBoer:2013gz} for a similar statement in the higher-spin AdS$_3$ case.} to obtain the thermal entropy is, thus, by first diagonalizing $a_\varphi$ and then using the diagonalized version of the connection denoted by $\lambda_\varphi$ to obtain the thermal entropy as
    \begin{equation}
        S_{\textrm{Th}}=2\pi k\left\langle\Lt_0\lambda_\varphi\right\rangle+\pi\kappa\gamma\left\langle\St c_\varphi\right\rangle,
    \end{equation}
or written in a more suggestive way as
    \begin{equation}
        S_{\textrm{Th}}=2\pi\left( -iM^v\left\langle\St c_\varphi\right\rangle-\left(-J^{\textrm v}-\frac{\left(M^{\textrm v}\right)^2}{\kappa}\right)\left\langle\Lt_0\lambda_\varphi\right\rangle\right).
    \end{equation}
\section{Metric interpretation}\label{Sec:Metric}

Up until this point, our analysis is based on the Chern-Simons formulation. As in the case of the higher-spin theory in AdS$_3$, we have imposed several conditions. Even though the matching of the asymptotic symmetries, the number of globally well defined gauge parameters and the thermal entropy fit the description of a spacelike warped black hole, it is hard to see 
at the level of the solution that the connection \eqref{eq:WAdSBCs} describes the desired black hole. In addition, there is the parameter $\gamma$ that we identified with the tilt of the cylinder where the dual WCFT is defined on. From a pure Chern-Simons perspective, this is a free and, up until now, undetermined parameter. However, one would expect that this parameter should be fixed in one way or the other by the geometry that is described by this Chern-Simons theory.  

In this section, we will provide a metric interpretation 
of our solution \eqref{eq:WAdSBCs} based on \cite{Hofman:2014loa}.
Our motivation is two-fold. The first is to explicitly verify that the connection \eqref{eq:WAdSBCs} describes a spacelike warped AdS$_3$ black hole. The second is to fix $\gamma$. The metric interpretation will also enable us to explicitly compute the Killing vectors as well as the inverse temperature and angular velocity, providing more evidence to support the claims made in Sec.~\ref{sec:EntropyMassAngularMomentumAndHolonmies}. 

\subsection{Mapping connection to metric}\label{sec:MapMetric}

A metric interpretation of the vacuum solutions given by the connection \eqref{eq:WAdSBCs} (with $\mu=0$ and $\nu=1$) has been first worked out in \cite{Hofman:2014loa}. In the following we will employ the same methods as in \cite{Hofman:2014loa} to show that the connection \eqref{eq:WAdSBCs} reproduces the metric of spacelike warped black holes. Thus, we will review briefly the main points of \cite{Hofman:2014loa} that are necessary to translate the connection \eqref{eq:WAdSBCs} into a metric form.

One of the points established in \cite{Hofman:2014loa}\footnote{Please note that in \cite{Hofman:2014loa} the Chern-Simons gauge field is labeled $B$ and the geometric variable $A$ in contrast to our notations.} is a precise relation between the gauge fields $\mathcal{A}$, $\mathcal{C}$ in the Chern-Simons formulation \eqref{eq:CSAction} and geometric variables $\mathcal{B}$ that encode the geometry. While the Chern-Simons fields are convenient for many purposes, one needs to determine the geometric variables from these fields in order to understand the geometry described by them. In the current setup, this is done as follows:
    \begin{itemize}
        \item Define three linearly independent vectors in $\mathfrak{sl}(2,\mathbb{R})$, ($\zeta_0^n,\zeta_1^n,\zeta_2^n$), and the inverse vectors ($\hat{\zeta}^0_n,\hat{\zeta}^1_n,\hat{\zeta}^2_n$) satisfying $\hat{\zeta}^I_n\zeta^n_J = \delta^I{}_J$ for $I,J=0,1,2$.
        Depending on the choice of these vectors, one can obtain either timelike, spacelike or null warped AdS$_3$ for vacuum solutions of \eqref{eq:WAdSBCs}.
        \item Then using these vectors one can determine the geometrical variables $\mathcal{B}$ via
        \begin{equation}
            \mathcal{B}^0=\sqrt{\left|\frac{8}{k\mathfrak{c}^2\alpha}\right|}\mathcal{C}-\frac{2\mathfrak{b}}{\mathfrak{c}}\hat{\zeta}^0_n\mathcal{A}^n,\,\mathcal{B}^1=\frac{\hat{\zeta}^1_n\mathcal{A}^n}{\sqrt{\mathfrak{c}}},\, \mathcal{B}^2=\frac{\hat{\zeta}^2_n\mathcal{A}^n}{\sqrt{\mathfrak{c}}}.
        \end{equation}
        The variable $\mathfrak{c}$ encodes the AdS radius, $\mathfrak{b}$ encodes the warping parameter and $\alpha$ is related to arbitrary rescalings of the time coordinate $t$.
        \item Using the $\mathfrak{sl}(2,\mathbb{R})\oplus\mathfrak{u}(1)$ invariant bilinear form $M_{IJ}=\zeta^n_I\eta_{nm}\zeta^m_J$ with $\eta_{nm}$ given in \eqref{eq:ToySLInvBilForm}, one can then determine the metric  $g_{\mu\nu}$ via
        \begin{equation}
            \extd s^2= g_{\mu\nu}dx^\mu dx^\nu =  \mathcal{B}^IM_{IJ}\mathcal{B}^J.
        \end{equation}
    \end{itemize}
Since our claim is that the connection \eqref{eq:WAdSBCs} describes spacelike warped AdS$_3$ black holes, we choose the following vectors in accordance with \cite{Hofman:2014loa}:
    \begin{equation}
        \zeta^0=(1,0,-1),\qquad \zeta^1=(1,0,1),\qquad\zeta^2=(0,1,0),
    \end{equation}
where the notation for the vectors is $\zeta=(+,0,-)$.
The metric obtained from the connection \eqref{eq:WAdSBCs} with $\mu=0$ and $\nu=1$ in this way is given by
    \begin{align}\label{eq:WAdSBHMetricOriginalCoords}
        \extd s^2=&\frac{\extd \rho^2}{2\mathfrak{c}}+\frac{16\extd t^2}{\mathfrak{c}^2k\alpha}+\frac{8\extd t\extd\varphi}{\mathfrak{c}^2}\left(\frac{16\pi\mathcal{K}}{k\alpha\kappa}-\sqrt{\frac{2}{k\alpha}}\mathfrak{b}\,\left(e^{\rho}+\mathfrak{L}e^{-\rho}\right)\right)+\frac{\extd\varphi^2}{2\mathfrak{c}^2}\Bigg(2\left(4\mathfrak{b}^2+\mathfrak{c}\right)\mathfrak{L}\nonumber\\
        &-\left(\mathfrak{c}-4\mathfrak{b}^2\right)\left(e^{2\rho}+\mathfrak{L}^2e^{-2\rho}\right)-64\pi\sqrt{\frac{2}{k\alpha}}\mathfrak{b}\frac{\mathcal{K}}{\kappa}\left(e^\rho+\mathfrak{L}e^{-\rho}\right)+\frac{512\pi^2\mathcal{K}^2}{k\alpha\kappa^2}\Bigg).
    \end{align}
This metric can be brought into a more familiar form by choosing a different radial coordinate $r$ via
    \begin{equation}
        \rho=2\log\left[\frac{\ell}{2}\sqrt{\mathfrak{c}}\left(\sqrt{r-r_+}+\sqrt{r-r_-}\right)\right],
    \end{equation}
where
    \begin{equation}
        \mathfrak{L} =
\frac{\ell^4\mathfrak{c}^2}{16}\left(r_+-r_-\right)^2,\qquad
        \mathcal{K} =  -\sqrt{\frac{k\alpha}{2}}\frac{\kappa}{16\pi}\ell^2\mathfrak{b}\mathfrak{c}\left(r_++r_--\sqrt{\frac{\mathfrak{c}r_+r_-}{\mathfrak{b}^2}}\right),
    \end{equation}
and $\ell$ denotes the radius of the warped AdS spacetime. If one in addition changes the sign of the angular coordinate $\varphi$ as $\varphi\rightarrow\varphi=-\phi$ and chooses the parameters $\mathfrak{b}$, 
$\mathfrak{c}$ and $\alpha$ as 
    \begin{equation}\label{eq:TMGWAdS3BHGeometricVariabels}
        \mathfrak{b}^2=\frac{\nu^2}{2\ell^2},\qquad\mathfrak{c}=\frac{\nu^2+3}{2\ell^2},\qquad\alpha=\frac{16}{\mathfrak{c}^2k\ell^2},
    \end{equation}
then one obtains precisely the metric of a rotating spacelike warped AdS$_3$ black hole \cite{Anninos:2008fx}
    \begin{align}\label{eq:WAdSBHMetricAndy}
        \frac{\extd s^2}{\ell^2}=&\extd t^2+\frac{\extd r^2}{\left(\nu^2+3\right)\left(r-r_+\right)\left(r-r_-\right)}+\left(2\nu r-\sqrt{r_+r_-\left(\nu^2+3\right)}\right)\extd t\extd\phi\nonumber\\
        &+\frac{r}{4}\left(3\left(\nu^2-1\right)r+\left(\nu^2+3\right)\left(r_++r_-\right)-4\nu\sqrt{r_+r_-\left(\nu^2+3\right)}\right)\extd\phi^2.
    \end{align}
In order to make the contact with the results derived previously using the Chern-Simons formulation a bit easier, however, we will use the metric \eqref{eq:WAdSBHMetricOriginalCoords} rather than \eqref{eq:WAdSBHMetricAndy} in the rest of this section.

\subsection{Killing vectors}\label{sec:KillingVectors}

The purpose of this section is to show that there are four Killing vectors of the metric \eqref{eq:WAdSBHMetricOriginalCoords} out of which two are globally well defined. In addition, we will show that the two globally well defined Killing vectors satisfy $\varepsilon=\xi^\mu\mathcal{A}_\mu$ and $\bar{\varepsilon}=\xi^\mu\mathcal{C}_\mu$. This justifies the requirement \eqref{eq:VariationMassAngMomFormula} for (the variation of) mass and angular momentum.

In order to find the Killing vectors of a spacetime given by a metric $g_{\mu\nu}$, one has to solve $\mathsterling_{\xi}g_{\mu\nu}=0$ for some vector field $\xi^\mu$. Solving this equation for the metric \eqref{eq:WAdSBHMetricOriginalCoords}, one obtains, as proclaimed, four linearly independent Killing vector fields:
    \begin{subequations}\label{eq:WAdSBHKillingVectors}
    \begin{align}
        \xi_1= & C_1e^{2\sqrt{\mathfrak{L}}\,\varphi}\left(\partial_\rho-\frac{\sqrt{2k\alpha}\mathfrak{b}\kappa\mathfrak{L}e^\rho-4\pi\mathcal{K}\left(e^{2\rho}+\mathfrak{L}\right)}{2\kappa\sqrt{\mathfrak{L}}\left(e^{2\rho}-\mathfrak{L}\right)}\partial_t-\frac{\left(e^{2\rho}+\mathfrak{L}\right)}{2\sqrt{\mathfrak{L}}\left(e^{2\rho}-\mathfrak{L}\right)}\partial_\varphi\right),\\
        \xi_2= & C_2e^{-2\sqrt{\mathfrak{L}}\,\varphi}\left(\partial_\rho+\frac{\sqrt{2k\alpha}\mathfrak{b}\kappa\mathfrak{L}e^\rho-4\pi\mathcal{K}\left(e^{2\rho}+\mathfrak{L}\right)}{2\kappa\sqrt{\mathfrak{L}}\left(e^{2\rho}-\mathfrak{L}\right)}\partial_t+\frac{\left(e^{2\rho}+\mathfrak{L}\right)}{2\sqrt{\mathfrak{L}}\left(e^{2\rho}-\mathfrak{L}\right)}\partial_\varphi\right),\\
        \xi_3= & C_3\partial_\varphi,\\
        \xi_4= & C_4\partial_t,
    \end{align}
    \end{subequations}
where $C_n$ are arbitrary integration constants. This is consistent with the result in 
\cite{Rooman:1998xf}. We note that the Killing 
vectors $\xi_3$ and $\xi_4$ are well defined both locally and globally.
The Killing vectors $\xi_1$ and $\xi_2$ on the other hand are not globally well defined for generic values of $\mathfrak{L}$ viz. $\mathcal{L}$ and $\mathcal{K}$, because of the periodicity of the angular coordinate $\varphi$. For the vacuum values \eqref{eq:CSVacuum}, however, these two Killing vectors become periodic. Thus, the vacuum state has four globally well defined Killing vectors as expected for warped AdS$_3$.

Having now both the explicit expression of the Killing vectors \eqref{eq:WAdSBHKillingVectors} as well as the expressions for the gauge parameters \eqref{eq:VacuumGaugeParameters} in the Chern-Simons formulation at hand, we can explicitly verify that we correctly identified mass and angular momentum previously in \eqref{eq:MassAndAngularMomentum}. This is simply done by taking the two Killing vectors $\xi_3$ and $\xi_4$. It is straightforward to check that they satisfy $\varepsilon=\xi^\mu\mathcal{A}_\mu$ and $\bar{\varepsilon}=\xi^\mu\mathcal{C}_\mu$ for the gauge parameters given by \eqref{eq:GaugeParameter} and \eqref{eq:VacuumGaugeParameters}. This result shows that our analysis combined with the metric interpretation proposed in \cite{Hofman:2014loa} fits with the usual analysis of the warped black holes in the metric formulation.

\subsection{Thermodynamic quantities}\label{sec:MetricThermodynamics}

In this part, we will first determine the location of the horizon for the metric \eqref{eq:WAdSBHMetricOriginalCoords} and then proceed in showing that the horizon is a Killing horizon. 
As a next step, we will determine angular velocity and surface gravity. In the end, we will show that these expressions exactly match \eqref{eq:InvTempAndAngPotPrelim} found in the Chern-Simons formulation, provided the parameter $\gamma$ is identified appropriately. This also gives us a way of fixing $\gamma$ completely.

First one needs to determine the location of the event horizon in the coordinates \eqref{eq:WAdSBHMetricOriginalCoords}. One particularly simple way of doing this is to find the constant-$\rho$ surface where the determinant of the induced metric changes sign, that is the location where the timelike Killing vector $\partial_t$ changes to be spacelike. On constant-$\rho$ slices the determinant of the induced metric $\gamma_{ij}$ reads
    \begin{equation}
        \det[\gamma_{ij}]=-\frac{8}{\mathfrak{c}^3k\alpha}\left(e^{\rho}-\mathfrak{L}e^{-\rho}\right)^2.
    \end{equation}
This expression is zero for
    \begin{equation}\label{eq:OriginalCoordinatedHorizon}
        \rho_\pm=\log\left[\pm\sqrt{\mathfrak{L}}\right].
    \end{equation}
We note that, in the coordinates chosen in \eqref{eq:WAdSBHMetricOriginalCoords}, only $\rho_+$ is admissible as the location of the horizon.

The angular velocity $\Omega$ can then be determined by demanding that the norm of the Killing vector $\xi=\partial_t+\Omega\partial_\varphi$ that generates the horizon vanishes at the horizon i.e. $\xi^\mu\xi_\mu\Big|_{\rho=\rho_+}=0$. This in turn means that the horizon located at \eqref{eq:OriginalCoordinatedHorizon} is a Killing horizon. Computing the norm of this Killing vector, one finds that the angular velocity is given by
    \begin{equation}
        \Omega= \frac{1}{\mathfrak{b}\sqrt{\frac{k\alpha}{2}}\sqrt{\mathfrak{L}}-\frac{4\pi\mathcal{K}}{\kappa}}.
    \end{equation}
This is exactly the same expression as \eqref{eq:InvTempAndAngPot} found previously using the Chern-Simons formulation upon identifying the parameter $\gamma$ as 
    \begin{equation}\label{eq:GammaIdentification}
        \gamma=\frac{\mathfrak{b}}{2}\sqrt{\frac{k\alpha}{2}}.
    \end{equation}
Since the horizon is a Killing horizon one can also associate a surface gravity and, thus, temperature to it. The surface gravity $\kappa_s$ can be determined straightforwardly via
    \begin{equation}
        \kappa_s=\sqrt{-\frac{1}{2}\nabla_\mu\xi_\nu\nabla^\mu\xi^\nu}\Big|_{\rho=\rho_+},
    \end{equation}
and for the metric \eqref{eq:WAdSBHMetricOriginalCoords} one obtains
    \begin{equation}
        \kappa_s=\frac{1}{\left(\frac{\mathfrak{b}}{2}\sqrt{\frac{k\alpha}{2}}-\frac{2\pi\mathcal{K}}{\kappa\sqrt{\mathfrak{L}}}\right)}.
    \end{equation}
Since temperature $T$ and surface gravity $\kappa_s$ are related via $T=\frac{\kappa_s}{2\pi}$, one finds for the inverse temperature
    \begin{equation}
        \beta=2\pi\left(\frac{\mathfrak{b}}{2}\sqrt{\frac{k\alpha}{2}}-\frac{2\pi\mathcal{K}}{\kappa\sqrt{\mathfrak{L}}}\right),
    \end{equation}
which again exactly coincides with \eqref{eq:InvTempAndAngPot} using the identification \eqref{eq:GammaIdentification}. Thus, the exact value of $\gamma$ depends on the translation from Chern-Simons to geometric variables. For the values \eqref{eq:TMGWAdS3BHGeometricVariabels} that yield the metric \eqref{eq:WAdSBHMetricAndy}, one obtains for example
    \begin{equation}
        \gamma=\frac{2\nu}{\nu^2+3}.
    \end{equation}
We note that, if one in addition chooses the Chern-Simons level $k$ as well as the $\mathfrak{u}(1)$ level $\kappa$ as\footnote{Please note that these expressions are given in units where $\ell=G=1$.}
    \begin{equation}
        k=\frac{5\nu^2+3}{6\nu\left(\nu^2+3\right)},\qquad\kappa=-\frac{\nu^2+3}{6\nu}\,, 
    \end{equation}
the thermal entropy \eqref{eq:CSEntropy} coincides with the entropy of spacelike warped AdS$_3$ black holes in topologically massive gravity (see e.g. Sec.~5 of \cite{Detournay:2012pc}).

\section{Conclusion and outlook}\label{conclusion_outlook}

In this paper we showed that lower-spin gravity described by a $\mathfrak{sl}(2,\mathbb{R})\oplus\mathfrak{u}(1)$ Chern-Simons theory contains solutions that can be interpreted as spacelike warped AdS$_3$ black holes. We argued that certain holonomy conditions give thermodynamically sensible relations between the canonical charges and the corresponding chemical potentials. The resulting thermal entropy is consistent with the first law of black hole thermodynamics and matches exactly the entropy of a WCFT at finite temperature. In order to support our claim, we also computed holographic entanglement entropy and found again perfect matching with what is expected from a WCFT at finite temperature, 
a holographic dual of the spacelike warped AdS$_3$ black hole. Furthermore, following the dictionary presented in \cite{Hofman:2014loa}, we provided a metric interpretation of our results. 
Our results show that a theory of lower-spin gravity provides a simple dual setup for a WCFT at finite temperature.

Our results presented in this paper can be extended in various ways. One possible extension would be to look for spacelike warped black hole solutions with additional higher-spin charges. Since our work shows that spacelike warped AdS$_3$ black holes can be described by using a Chern-Simons theory with $\mathfrak{sl}(2,\mathbb{R})\oplus\mathfrak{u}(1)$ gauge symmetry, it is suggestive to look at Chern-Simons theories with extended gauge symmetry which contains $\mathfrak{sl}(2,\mathbb{R})\oplus\mathfrak{u}(1)$ as a subalgebra.  The most simple possible example for such an extension might be the nonprincipal embedding of $\mathfrak{sl}(3,\mathbb{R})$. Even though this particular case would not be a higher-spin extension in the strict sense (since there are no excitations with spin-$s>2$), it would be interesting to look at such a theory as a first step towards more complicated examples. The results found in \cite{Breunhoelder:2015waa} for null warped AdS$_3$ in higher-spin gravity suggests that such an endeavor is indeed promising. In this work the authors found boundary conditions for a spin-3 gravity theory in AdS that asymptote to null warped AdS$_3$ and whose asymptotic symmetries are given by one copy of the Polyakov-Bershadsky $\mathcal{W}_3^{(2)}$ algebra \cite{Polyakov:1989dm,Bershadsky:1990bg}. Since there exist various examples\footnote{For AdS$_3$ higher-spin theories see e.g. \cite{Campoleoni:2011hg,Bunster:2014mua}. Another example with non-AdS asymptotics, namely Lobachevsky can be found for example in \cite{Riegler:2012fa,Afshar:2012nk}.} of consistent boundary conditions in (non-)AdS$_3$ higher-spin gravity theories that involve the nonprincipal embedding of $\mathfrak{sl}(3,\mathbb{R})$ and the $\mathcal{W}_3^{(2)}$ algebra, it seems plausible that $\mathcal{W}_3^{(2)}$ algebra will make its appearance as the asymptotic symmetry algebra of a ``higher''-spin extended spacelike warped black hole.

Since lower-spin gravity provides a simple dual model for a WCFT, it would be very interesting to consider a supersymmetric extension and then apply the localization techniques used in \cite{Iizuka:2015jma,Honda:2015hfa,Honda:2015mel} to compute the full partition function of lower-spin gravity. This could potentially lead to many new insights regarding WCFTs and quantum gravity in general.

Another interesting direction is related to soft excitations of black hole horizons in three dimensions. Starting with the work \cite{Afshar:2016wfy}, consistent near-horizon boundary conditions for BTZ black holes were found that lead to a very simple near-horizon symmetry algebra, namely two affine $\hat{\mathfrak{u}}(1)$ current algebras. The thermal entropy of the BTZ black hole expressed in these near-horizon variables takes a strikingly simple form
    \begin{equation}\label{eq:HairyEntropyConclusion}
        S_{\textrm{Th}}=2\pi\left(\Jt_0+\bar{\Jt}_0\right),
    \end{equation}
where $\Jt_0$ and $\bar{\Jt}_0$ denote the zero modes of the affine $\hat{\mathfrak{u}}(1)$ current algebras. Following up on this work there were many checks in different setups \cite{Afshar:2016wfy,Setare:2016vhy,Afshar:2016uax,Grumiller:2016kcp,Afshar:2016kjj,Ammon:2017vwt,Gonzalez:2017sfq,Grumiller:2017otl,Setare:2017xlu} as to how general this result of the entropy is in three-dimensional gravity. Since lower-spin gravity provides a new theory to test the generality of the entropy formula \eqref{eq:HairyEntropyConclusion} it might be interesting to see if one can find consistent near-horizon boundary conditions that either confirm or contradict \eqref{eq:HairyEntropyConclusion} in this setup.

\subsection*{Acknowledgments}

We would like to thank Alain Buisseret for collaboration at the early stages of this work \cite{AlainThesis} as well as Luis Apolo, Glenn Barnich, Alejandra Castro, Diego Hofman and Wei Song for valuable discussions. M. R. wants to thank Martin Ammon and the TPI at the Friedrich-Schiller-Universit{\"a}t Jena for the opportunity of an extended visit during the early stages of this project. 
T. A. and S. D. are supported in part by the ARC grant ``Holography, Gauge Theories and Quantum Gravity Building Models of Quantum Black Holes''. 
S. D. is a Research Associate of the Fonds de la Recherche Scientifique F.R.S.-FNRS (Belgium). He is also supported by IISN-Belgium (convention 4.4503.15) and benefited from the support of the Solvay Family. The research of M. R. is supported by the ERC Starting Grant No. 335146 ``HoloBHC".

\begin{appendix}

\section{Thermodynamics of BTZ black holes}\label{sec:AppendixA}

Most of the requirements that we made in the beginning of this work in the Chern-Simons formulation are inspired by how the thermodynamics of BTZ black holes are described in SL$(2, \mathbb{R})\times$ SL$(2, \mathbb{R})$ Chern-Simons theory. Hence we give a brief review\footnote{For the interested reader a (nonexhaustive) list of reviews and further reading regarding the relations between three-dimensional (higher-spin) gravity and Chern-Simons theories is given by \cite{Witten:1988hc,Gutperle:2011kf,Ammon:2011nk,Campoleoni:2012hp,deBoer:2013gz,Gomez:2013sfb,Bunster:2014mua,Donnay:2016iyk,Riegler:2017fqv,Prohazka:2017lqb}.} thereof in this appendix, focusing on the relevant points for our work. We start with a very brief review of the main points of thermodynamics for BTZ black holes in the usual metric formulation of Einstein gravity in Appendix~\ref{sec:AppendixA1} and then continue in Appendix~\ref{sec:AppendixA2} to review how these geometric statements translate into the Chern-Simons formulation.

\subsection{BTZ black boles in the metric formulation}\label{sec:AppendixA1}

The metric of the BTZ black hole \cite{Banados:1992wn,Banados:1992gq} is given by
    \begin{equation}\label{eq:BTZBlackHoleCoords}
        \extd s^2=-N^2\extd t^2+N^{-2}\extd r^2+r^2\left(\extd\varphi+N^\varphi\extd t\right)^2,
    \end{equation}
with
    \begin{equation}
        N^2=\frac{\left(r^2-r_+^2\right)\left(r^2-r_-^2\right)}{r^2\ell^2},\qquad N^\varphi=\frac{r_+r_-}{r^2\ell}.
    \end{equation}
Here the spatial coordinates 
$r, \varphi$ take values in $0\leq r<\infty$, $\varphi\in[0,2\pi]$ and the temporal one $t$ is in 
$-\infty<t<\infty$. 
This metric has two horizons $r_\pm$, that are given in terms of mass $M$, angular momentum $J$ and AdS radius $\ell$ as
    \begin{equation}
        r_\pm=\ell\left(\sqrt{\frac{1}{2k}\left(\ell M-J\right)}\pm\sqrt{\frac{1}{2k}\left(\ell M+J\right)}\right).
    \end{equation}
Using a different kind of radial coordinate $\rho$ that is defined via     \begin{equation}\label{eq:RelationRhoRBTZ}
        \rho=\log\left[\frac{1}{2\ell}\left(\sqrt{r^2-r_+^2}+\sqrt{r^2-r_-^2}\right)\right],
    \end{equation}
and in addition introducing the parameters $\mathcal{L}$ and $\bar{\mathcal{L}}$ as
    \begin{equation}\label{eq:BTZMassAngularMomentumRelationCharges}
       \ell M = 2\pi\left(\mathcal{L}+\bar{\mathcal{L}}\right),\qquad J = -2\pi\left(\mathcal{L}-\bar{\mathcal{L}}\right),
    \end{equation}
one can bring the metric \eqref{eq:BTZBlackHoleCoords} into Fefferman-Graham form\footnote{See e.g. \cite{Papadimitriou:2004ap} for a review.} as 
	\begin{align}\label{eq:AdS3MetricLLbar}
		\extd s^2=&\ell^2\left[\extd\rho^2+\frac{2\pi}{k}\left(\mathcal{L}(\extd x^+)^2+\bar{\mathcal{L}}(\extd x^-)^2\right)\right.\nonumber\\
		&\left.-\left(e^{2\rho}+\frac{4\pi^2}{k^2}\mathcal{L}\bar{\mathcal{L}}e^{-2\rho}\right)\extd x^+\extd x^-\right]. 
	\end{align}
Here we used light-cone coordinates $x^\pm=\frac{t}{\ell}\pm\varphi$ for convenience. Having the metric in a form such as \eqref{eq:AdS3MetricLLbar} makes comparison with the Chern-Simons formulation easy and, thus, we will continue to use this form.

Before proceeding to the Chern-Simons formulation, it is useful to first determine some essential quantities characterizing the black hole in the metric formulation. The first important ingredients are the Killing vectors of the metric \eqref{eq:AdS3MetricLLbar}. It is straightforward to verify that the Killing vectors of \eqref{eq:AdS3MetricLLbar} are given by
    \begin{subequations}\label{eq:BTZKillingVectors}
    \begin{align}
        \xi_1 = & C_1\left(-\frac{e^{2x^+\sqrt{\mathfrak{L}}}}{2}\partial_\rho+\frac{(e^{4\rho}+\mathfrak{L}\bar{\mathfrak{L}})e^{2x^+\sqrt{\mathfrak{L}}}}{2(e^{4\rho}-\mathfrak{L}\bar{\mathfrak{L}})\sqrt{\mathfrak{L}}}\partial_{x^+}+\frac{e^{2x^+\sqrt{\mathfrak{L}}+2\rho}\sqrt{\mathfrak{L}}}{(e^{4\rho}-\mathfrak{L}\bar{\mathfrak{L}})}\partial_{x^-}\right),\\
        \xi_2 = & C_2\left(-\frac{e^{-2x^+\sqrt{\mathfrak{L}}}}{2}\partial_\rho-\frac{(e^{4\rho}+\mathfrak{L}\bar{\mathfrak{L}})e^{-2x^+\sqrt{\mathfrak{L}}}}{2(e^{4\rho}-\mathfrak{L}\bar{\mathfrak{L}})\sqrt{\mathfrak{L}}}\partial_{x^+}-\frac{e^{-2x^+\sqrt{\mathfrak{L}}+2\rho}\sqrt{\mathfrak{L}}}{(e^{4\rho}-\mathfrak{L}\bar{\mathfrak{L}})}\partial_{x^-}\right),\\
        \xi_3 = & C_3\partial_{x^+},\\
        \xi_4 = & C_4\left(-\frac{e^{2x^-\sqrt{\bar{\mathfrak{L}}}}}{2}\partial_\rho+\frac{e^{2x^-\sqrt{\bar{\mathfrak{L}}}+2\rho}\sqrt{\bar{\mathfrak{L}}}}{(e^{4\rho}-\mathfrak{L}\bar{\mathfrak{L}})}\partial_{x^+}+\frac{(e^{4\rho}+\mathfrak{L}\bar{\mathfrak{L}})e^{2x^-\sqrt{\bar{\mathfrak{L}}}}}{2(e^{4\rho}-\mathfrak{L}\bar{\mathfrak{L}})\sqrt{\bar{\mathfrak{L}}}}\partial_{x^-}\right),\\
        \xi_5 = & C_5\left(-\frac{e^{-2x^-\sqrt{\bar{\mathfrak{L}}}}}{2}\partial_\rho-\frac{e^{-2x^-
        \sqrt{\bar{\mathfrak{L}}}+2\rho}\sqrt{\bar{\mathfrak{L}}}}{(e^{4\rho}-\mathfrak{L}\bar{\mathfrak{L}})}\partial_{x^+}-\frac{(e^{4\rho}+\mathfrak{L}\bar{\mathfrak{L}})e^{-2x^-\sqrt{\bar{\mathfrak{L}}}}}{2(e^{4\rho}-\mathfrak{L}\bar{\mathfrak{L}})\sqrt{\bar{\mathfrak{L}}}}\partial_{x^-}\right),\\
        \xi_6 = & C_6\partial_{x^-},
    \end{align}
    \end{subequations}
where we used the abbreviations $\mathfrak{L}\equiv\frac{2\pi}{k}\mathcal{L}$ and $\bar{\mathfrak{L}}\equiv\frac{2\pi}{k}\bar{\mathcal{L}}$. Because of the periodicity of the angular coordinate $\varphi$, 
for generic values of
$\mathfrak{L}$ and 
$\bar{\mathfrak{L}}$, 
only two out of these six Killing vectors are globally well defined i.e. $\xi_3$ and $\xi_6$. The other four Killing vectors $\xi_1$, $\xi_2$, $\xi_4$ and $\xi_5$ become globally well defined once $\mathfrak{L}$ and $\bar{\mathfrak{L}}$ become negative. Looking at the resulting spacetimes, however, one finds that, for almost all possible combinations of mass and angular momentum satisfying this requirement, the resulting spacetime exhibits pathologies such as closed timelike curves, naked singularities or angular excesses/defects with one important exception. That is, the case with $\ell M=-\frac{k}{2}$ and $J=0$ that yields global AdS$_3$. Thus, global AdS$_3$ can be interpreted as the vacuum state of the BTZ black hole.

In order to determine the entropy of the BTZ black hole one first has to determine the location of the outer horizon in the coordinates used in \eqref{eq:AdS3MetricLLbar}. This can be done by either reading the location off from \eqref{eq:RelationRhoRBTZ} or alternatively by looking at the value of $\rho$ where the induced metric of \eqref{eq:AdS3MetricLLbar} on slices of constant $\rho$ vanishes. Both yield the same result, namely
    \begin{equation}
        \rho_{\textrm{H}}=\frac{1}{4}\log\left[\frac{4\pi^2}{k^2}\mathcal{L}\bar{\mathcal{L}}\right].
    \end{equation}
Once the position of the horizon is located, it is straightforward to determine the angular velocity $\Omega$ by requiring that the norm of the Killing vector $\xi=\ell\partial_t+\Omega\partial_\varphi$ vanishes at the horizon i.e.
    \begin{equation}
        \xi^\mu\xi_\mu\Big|_{\rho=\rho_{\textrm{H}}}=0.
    \end{equation}
Subsequently one can also determine the surface gravity $\kappa_s$ of the horizon and, thus, also the temperature $T=\frac{\kappa_s}{2\pi}=\beta^{-1}$ via
    \begin{equation}
        \kappa_s=\sqrt{-\frac{1}{2}\nabla_\mu\xi_\nu\nabla^\mu\xi^\nu}\Big|_{\rho=\rho_{\textrm{H}}}.
    \end{equation}
By following this, for the metric \eqref{eq:AdS3MetricLLbar},
the explicit form of $\beta$ and
$\Omega$ turns out to be 
    \begin{equation}\label{eq:BTZInverseTemperatureAndAngularPotential}
        \beta=\sqrt{\frac{k\pi}{8}}\frac{\sqrt{\mathcal{L}}+\sqrt{\bar{\mathcal{L}}}}{\sqrt{\mathcal{L\bar{\mathcal{L}}}}},\qquad\Omega=-\frac{\sqrt{\mathcal{L}}-\sqrt{\bar{\mathcal{L}}}}{\sqrt{\mathcal{L}}+\sqrt{\bar{\mathcal{L}}}}.
    \end{equation}
The thermal entropy can then be determined via the area law $S_{\textrm{Th}}=\frac{A}{4G}$ and one obtains
    \begin{equation}\label{eq:BTZThermalEntropy}
        S_{\textrm{Th}}=2\pi\left(\sqrt{2\pi k \mathfrak{L}}+\sqrt{2\pi k \bar{\mathfrak{L}}}\right).
    \end{equation}

\subsection{BTZ black holes in the Chern-Simons formulation}\label{sec:AppendixA2}

Einstein gravity with negative cosmological constant can be described by the difference of two Chern-Simons actions \cite{Achucarro:1987vz,Witten:1988hc}
	\begin{equation}\label{eq:ChernSimonsActionAAbar}
		I_{\textnormal{EH}}^{\textnormal{AdS}}=I_{\textnormal{CS}}[A]-I_{\textnormal{CS}}[\bar{A}],
	\end{equation}
where
	\begin{equation}\label{eq:ChernSimonsAction}
		I_{\textnormal{CS}}[A]=\frac{k}{4\pi}\int_{\mathcal{M}}\left<A\wedge\extd A+\frac{2}{3}A\wedge A\wedge A\right>.
	\end{equation}
The Chern-Simons level $k$, the AdS radius $\ell$ and Newton's constant $G$ are related via $k=\frac{\ell}{4G}$. The Chern-Simons connections $A$ and $\bar{A}$ both take values in $\mathfrak{sl}(2,\mathbb{R})$. We choose the basis as
    \begin{equation}
        [\Lt_n,\Lt_m] = (n-m)\Lt_{n+m},
    \end{equation}
for $m, n=0,\pm1$ whose invariant bilinear form $\langle\ldots\rangle$ is give in \eqref{eq:ToySLInvBilForm}. The manifold $\mathcal{M}$ is given by a cylinder with coordinates $0\leq\rho<\infty$, $-\infty<t<\infty$ and $\varphi\in[0,2\pi]$.

The BTZ black hole \cite{Banados:1992wn,Banados:1992gq} in this setup is described by the following connection:
    \begin{subequations}\label{eq:BTZConnectionsAAbar}
    \begin{align}
        A = & b^{-1}(a+d)b,\qquad \bar{A}=b(\bar{a}+d)b^{-1},\\
        a = & (\Lt_1-\frac{2\pi}{k}\mathcal{L}\Lt_{-1})(\extd\varphi+\frac{\extd t}{\ell}),\\
        \bar{a} = & (\Lt_{-1}-\frac{2\pi}{k}\bar{\mathcal{L}}\Lt_{1})(\extd\varphi-\frac{\extd t}{\ell}),
    \end{align}
    \end{subequations}
where a popular choice\footnote{One of the reasons why this choice of gauge is favored is that this naturally leads to the metric \eqref{eq:AdS3MetricLLbar}.} for $b$ is $b=e^{\rho\Lt_0}$ and $\mathcal{L}$ and $\bar{\mathcal{L}}$ are the constants that encode mass $M$ and angular momentum $J$ of the BTZ black hole that we introduced in \eqref{eq:BTZMassAngularMomentumRelationCharges}. The metric in this formulation is recovered via $g_{\mu\nu}=\frac{\ell^2}{2}\left\langle(A_\mu-\bar{A}_\mu)(A_\nu-\bar{A}_\nu)\right\rangle$ and yields \eqref{eq:AdS3MetricLLbar}. The metric in the original BTZ coordinates \eqref{eq:BTZBlackHoleCoords} can be recovered by simply replacing $\rho$ with the radial coordinate $r$ according to \eqref{eq:RelationRhoRBTZ}.

In the metric formulation all thermodynamic quantities have a clear and geometric interpretation. In the following we show how these geometric statements translate into the Chern-Simons formulation. Now we consider the following questions in the Chern-Simons formulation:
\begin{itemize}
    \item How to identify the vacuum state?
    \item How to relate inverse temperature $\beta$ and angular velocity $\Omega$ with the charges $\mathcal{L}$ and $\bar{\mathcal{L}}$?
    \item How to determine the thermal entropy?
\end{itemize}
The main idea to tackle these questions in the Chern-Simons formulation is basically to look at the holonomies of the gauge connections $A$ and $\bar{A}$. These holonomies around given cycles should be trivial, which is tantamount to requiring that these cycles can be contracted in a smooth manner. 
\subsubsection{The vacuum state}
Looking at the holonomies around the $\varphi$-cycle that is
    \begin{equation}\label{eq:BTZHolonomiesVacuumConditions}
        \textrm{Hol}_\varphi(A)=b^{-1}e^{\oint a_\varphi}b,\qquad \textrm{Hol}_\varphi(\bar{A})=be^{\oint \bar{a}_\varphi}b^{-1},
    \end{equation}
one can see that these holonomies -- that are nontrivial for generic values of $\mathcal{L}$ and $\bar{\mathcal{L}}$ -- become trivial i.e. $\textrm{Hol}_\varphi(A)=\textrm{Hol}_\varphi(\bar{A})=-\unity$ for the following values of $\mathcal{L}$ and $\bar{\mathcal{L}}$:
    \begin{equation}\label{eq:BTZHolonomiesVacuum}
        \mathcal{L}=\bar{\mathcal{L}}=-\frac{k}{8\pi}\left(1+2n\right)^2,
    \end{equation}
where $n$ is some integer number. For $n=0$ this corresponds exactly to $\ell M=-\frac{k}{2}$ and $J=0$, 
i.e. global AdS$_3$ in terms of a metric interpretation. Thus, these holonomies can serve as a tool to determine the vacuum state.

Alternatively one can also make use of the fact that diffeomorphisms or in other words Killing vectors $\xi^\mu$ in the metric formulation are on-shell equivalent to gauge transformations of the form
    \begin{equation}
        \delta_\epsilon A_\mu=\partial_\mu\epsilon+[A_\mu,\epsilon],\qquad\delta_{\bar{\epsilon}} \bar{A}_\mu=\partial_\mu\bar{\epsilon}+[\bar{A}_\mu,\bar{\epsilon}],
    \end{equation}
with gauge parameters $\epsilon=\xi^\mu A_\mu$ and $\bar{\epsilon}=\xi^\mu\bar{A}_\mu$. One can show that the Killing vectors \eqref{eq:BTZKillingVectors} in combination with the connections \eqref{eq:BTZConnectionsAAbar} yields six linearly-independent gauge parameters $\epsilon_a$ and $\bar{\epsilon}_b$ with $a,b=1,2,3$ satisfying $\delta_\epsilon A_\mu=\delta_{\bar{\epsilon}}\bar{A}_\mu=0$. Out of these six gauge parameters, again two are globally well defined and the other four become globally well defined only for the vacuum values \eqref{eq:BTZHolonomiesVacuum}. This corresponds to the symmetry enhancement that we have encountered at the level of the Killing vectors in the metric formulation. 
\subsubsection{Inverse temperature and angular velocity}
A similar approach based on the holonomies can be taken for computing the thermal entropy of the BTZ black hole in the Chern-Simons formulation. This is usually done by first performing a Wick rotation of the Lorentzian time coordinate $t$ to Euclidean time $t_{\textrm{E}}$ which is then compactified. This changes the topology of the solid cylinder to that of a solid torus. One can introduce complex coordinates $(z,\bar{z})$ by analytically continuing the light-cone directions $x^+\rightarrow\frac{it_{\textrm{E}}}{\ell}+\varphi = z$ and $x^-\rightarrow\frac{it_{\textrm{E}}}{\ell}-\varphi =-\bar{z}$. These coordinates are then identified as
    \begin{equation}
        z\simeq z+2\pi\simeq z+2\pi\tau,\qquad \bar{z}\simeq \bar{z}+2\pi\simeq \bar{z}+2\pi\bar{\tau},
    \end{equation}
where $\tau$ is the modular parameter of the boundary torus. Following e.g. \cite{Gutperle:2011kf,Ammon:2011nk} (see also \cite{deBoer:2013gz} or \cite{Bunster:2014mua}) smooth black hole solutions are defined via the holonomies of the contractible cycles around the constant-$\rho$ torus. This is tantamount to the statement that the holonomies belong to the center of the gauge group \cite{Castro:2011fm,Castro:2011iw}. For rotating BTZ black holes these holonomies are given by
    \begin{equation}
        \textrm{Hol}_{\tau,\bar{\tau}}(A)=b^{-1}e^\omega b,\qquad\textrm{Hol}_{\tau,\bar{\tau}}(\bar{A})=be^{\bar{\omega}}b^{-1},
    \end{equation}
where
    \begin{equation}\label{eq:BTZHolonomies}
        \omega=2\pi(\tau a_z+\bar{\tau}a_{\bar{z}}),\qquad \bar{\omega}=2\pi(\tau\bar{a}_z+\bar{\tau}\bar{a}_{\bar{z}}).
    \end{equation}
The requirement to have trivial holonomies is a restriction on the eigenvalues of \eqref{eq:BTZHolonomies} namely
    \begin{equation}
        \textrm{Eigen}[\omega]=\textrm{Eigen}[\bar{\omega}]=\textrm{Eigen}[2\pi i \Lt_0].
    \end{equation}
The quantities $\omega$, $\bar{\omega}$
can also be written in terms of the Lorentzian connections $a_t=a_z-a_{\bar{z}}$ and $a_\varphi=a_z+a_{\bar{z}}$ ($\bar{a}_t=\bar{a}_z-\bar{a}_{\bar{z}}$ and $\bar{a}_\varphi=\bar{a}_z+\bar{a}_{\bar{z}}$). In addition one can use the fact that the modular parameters and the thermodynamic potentials for the BTZ black hole have to satisfy\footnote{This relation can be determined from the metric side where this statement translates to the absence of a conical singularity at the horizon of the BTZ black hole.}
    \begin{equation}\label{eq:BTZModularParameters}
        \tau=\frac{i\beta}{2\pi}(1+\Omega),\qquad \bar{\tau}=-\frac{i\beta}{2\pi}(1-\Omega).
    \end{equation}
Using this one can write the relations \eqref{eq:BTZHolonomies} equivalently as
    \begin{equation}\label{eq:BTZHolonomiesLorentzian}
        \omega=i\beta(a_t+\Omega a_\varphi)\equiv i h,\, \bar{\omega}=i\beta(\bar{a}_t+\Omega \bar{a}_\varphi)\equiv i \bar{h}.
    \end{equation}
Thus, one can also say that 
the holonomy conditions \eqref{eq:BTZHolonomies} 
give restrictions on the eigenvalues of $h$ and $\bar{h}$ such that
    \begin{equation}
        \textrm{Eigen}[h]=\textrm{Eigen}[\bar{h}]=\textrm{Eigen}[2\pi \Lt_0].
    \end{equation}
In the absence of rotation $\Omega=0$, this means that the Euclidean time cycle has to be contractible and, thus, also the holonomy around that cycle has to be trivial. For the rotating solutions, the contractible cycle is a more general thermal cycle given by \eqref{eq:BTZHolonomies} (or equivalently \eqref{eq:BTZHolonomiesLorentzian}).

\subsubsection{Thermal entropy and first law}
These holonomy conditions give precise relations between the thermodynamic potentials $\beta$, $\Omega$ and the charges $\mathcal{L}$ and $\bar{\mathcal{L}}$. Once these relations are fixed, there are various ways of computing the thermal entropy from the connections \eqref{eq:BTZConnectionsAAbar} (see e.g. \cite{deBoer:2013gz}). One particular simple way is via integrating the first law of black hole thermodynamics
    \begin{equation}\label{eq:BTZFirstLaw}
        \delta S_{\textrm{Th}}=\beta \left(\delta M-\Omega\delta J\right).
    \end{equation}
In order to be able to integrate this first law, one needs to know how mass $M$ and angular momentum $J$ are related to the constants $\mathcal{L}$ and $\bar{\mathcal{L}}$. In \eqref{eq:BTZMassAngularMomentumRelationCharges}
we have already presented this relation without 
any explanation. We will explain this point here. 
In the metric formulation mass and angular momentum are the canonical boundary charges associated to translations in time as well as in the angular direction i.e. the Killing vectors $\partial_t$ and $-\partial_\varphi$. The (variation of the) canonical boundary charges in the Chern-Simons formulation is given by
    \begin{equation}
        \delta Q[\epsilon]+\delta\bar{Q}[\bar{\epsilon}]=\frac{k}{2\pi}\int\extd\varphi\left\langle\epsilon\delta a_\varphi\right\rangle-\frac{k}{2\pi}\int\extd\varphi\left\langle\bar{\epsilon}\delta \bar{a}_\varphi\right\rangle,
    \end{equation}
where $\epsilon$ and $\bar{\epsilon}$ are gauge parameters that preserve the boundary boundary conditions via $\delta_\epsilon A_\mu=\partial_\mu\epsilon+[A_\mu,\epsilon]$ and $\delta_{\bar{\epsilon}} \bar{A}_\mu=\partial_\mu\bar{\epsilon}+[\bar{A}_\mu,\bar{\epsilon}]$. The link between the charges in the metric formulation and the ones in the Chern-Simons formulation is connected with the on-shell relation $\epsilon=\xi^\mu A_\mu$, $\bar{\epsilon}=\xi^\mu\bar{A}_\mu$ between gauge parameters $\epsilon$, $\bar{\epsilon}$ and Killing vectors $\xi^\mu$. Using this relation one finds that the variation of the mass and angular momentum in the Chern-Simons formulation is given by
    \begin{subequations}
    \begin{align}
        \ell\delta M & = \delta Q[\epsilon\big|_{\partial_t}]+\delta\bar{Q}[\bar{\epsilon}\big|_{\partial_t}] =\frac{k}{2\pi}\int\extd\varphi\left\langle a_t\delta a_\varphi\right\rangle-\frac{k}{2\pi}\int\extd\varphi\left\langle \bar{a}_t\delta \bar{a}_\varphi\right\rangle,\\
        \delta J & = \delta Q[\epsilon\big|_{-\partial_\varphi}]+\delta\bar{Q}[\bar{\epsilon}\big|_{-\partial_\varphi}]=-\frac{k}{2\pi}\int\extd\varphi\left\langle a_\varphi\delta a_\varphi\right\rangle+\frac{k}{2\pi}\int\extd\varphi\left\langle \bar{a}_\varphi\delta \bar{a}_\varphi\right\rangle.
    \end{align}
    \end{subequations}
By integrating these expressions one exactly finds the relations \eqref{eq:BTZMassAngularMomentumRelationCharges}.

Once mass, angular momentum as well as the corresponding chemical potentials are properly determined, one can proceed in integrating the first law of thermodynamics \eqref{eq:BTZFirstLaw} to obtain the thermal entropy. For the connection \eqref{eq:BTZConnectionsAAbar} one obtains exactly the same expressions for $\beta$ and $\Omega$ as in \eqref{eq:BTZInverseTemperatureAndAngularPotential}. This then leads to the thermal entropy \eqref{eq:BTZThermalEntropy}.    
\end{appendix}

\bibliographystyle{fullsort}
\bibliography{Bibliography}

\end{document}